\documentclass[12pt]{article}

\usepackage{graphics}
\usepackage{times}
\usepackage{amsmath}
\usepackage{amssymb}
\usepackage{aaspp4} 

\newcommand{\diff}{\mathrm{d}}

\newcommand{\Var}{\mathrm{Var}}
\newcommand{\Cov}{\mathrm{Cov}}
\newcommand{\func}[1]{\mathop{\textrm{#1}}\nolimits}
\renewcommand{\vec}[1]{\mbox{\boldmath $#1$\unboldmath}}

\slugcomment{Submitted to \textit{The Astrophysical Journal}.} 
\lefthead{Bertin \& Lombardi}
\righthead{Double Lenses}
\begin{document}

   \title{Double lenses}
   \author{Giuseppe Bertin \and Marco Lombardi\altaffilmark{1}}
   \affil{Scuola Normale Superiore,
     Piazza dei Cavalieri 7, I 56126 Pisa, Italy}
   \altaffiltext{1}{European Southern Observatory,
     Karl-Schwarzschild-Stra\ss e 2, Garching bei M\"unchen, D-85748,
     Germany}


   \begin{abstract}
     
     The analysis of the shear induced by a single cluster, acting as
     a gravitational lens, on the images of a large number of
     background galaxies is all centered around the curl-free
     character of a well-known vector field that can be derived from
     the data; in particular, the mass reconstruction methods,
     currently producing many interesting astrophysical applications,
     are all based on this tensorial property of the induced shear.
     Such basic property breaks down when the source galaxies happen
     to be observed through two clusters at different redshifts,
     partially aligned along the line of sight, because an asymmetric
     part in the Jacobian matrix associated with the ray tracing
     transformation is expected, now that the incoming light rays are
     bent twice. In this paper we address the study of double lenses
     and obtain five main results. (i) First we generalize the
     procedure to extract the available information, contained in the
     observed shear field, from the case of a single lens to that of a
     double lens. (ii) Then we evaluate the possibility of detecting
     the signature of double lensing given the known properties of the
     distribution of clusters of galaxies. In particular, we show that
     a few configurations are likely to be present in the sky, for
     which the small effects characteristic of double lensing may
     already be within detection limits (i.e., we show that the signal
     involved is expected to be larger than what could be produced by
     statistical noise, which includes effects associated with the
     distribution of the source ellipticities and with the spread in
     redshift of the lensed galaxies). (iii) As a different
     astrophysical application, we demonstrate how the method can be
     used to detect the presence of a dark cluster that might happen
     to be partially aligned with a bright cluster studied in terms of
     statistical lensing; if the properties of the bright cluster are
     well constrained by independent diagnostics, the location and the
     structure of the dark cluster can be reconstructed. (iv) In
     addition, we show that the redshift distribution of the source
     galaxies, which in principle might also contribute to break the
     curl-free character of the shear field, actually produces
     systematic effects typically two orders of magnitude smaller than
     the double lensing effects we are focusing on. (v) Remarkably, a
     discussion of relevant contributions to the noise of the shear
     measurement has brought up an intrinsic limitation of weak
     lensing analyses, since one specific contribution, associated
     with the presence of a non-vanishing two-galaxy correlation
     function, turns out not to decrease with the density of source
     galaxies (and thus with the depth of the observations). In our
     mathematical framework, the contribution of the small asymmetry
     in the Jacobian matrix induced by double lensing is retained
     consistently up to two orders in the weak lensing asymptotic
     expansion. The analysis is then checked and exemplified by means
     of simulations.

   \end{abstract}
   


\section{Introduction}

The study of statistical gravitational lensing effects on large
numbers of source galaxies due to intervening matter has been a
subject of major and rapid progress both from the methodological and
from the observational point of view (e.g., see Kaiser \& Squires
1993; Luppino \& Kaiser 1997; Taylor et al.\ 1998; Lombardi \& Bertin
1998a, b, hereafter Paper~I, II). This indeed offers a very promising
approach to study the overall mass distribution, especially on the
scale of clusters of galaxies. In fact, the most appealing aspect of
this line of research is its diagnostic potential, with the
possibility to measure mass distributions independently of the
traditional tools that have to rely on the use of dynamical models.

The interest in this method has gained further momentum from the most
recent developments of telescopes and instrumentation. Very deep
images are now bringing in accurate data that allow us to analyze the
shape of thousands of galaxies, with a density of nearly $200 \mbox{
  galaxies arcmin}^{-2}$ already reached (see Hoekstra et al.\ 2000).
The use of information from so many sources makes it possible to
measure very small lensing effects, which would have been beyond
expectation just a few years ago.

The work in this area is thus mostly addressing two directions. On the
methodological side, the main hope is to develop mass reconstruction
algorithms that are simple, flexible, and reliable, with full control
of the errors involved in the procedure that goes from the shear data
to the inferred properties of the lens (see, e.g., Bartelmann et al.
1996; Seitz \& Schneider 1997; Paper~I; Paper~II). On the side of data
analysis, a variety of important issues have to be properly faced and
addressed, in order to secure accurate shape measurements as free as
possible from all the undesired or spurious distortions that are
associated with the instrument and with the observation conditions
(Kaiser, Squires, \& Broadhurst 1995).

Within this general pespective, the present paper belongs to a third
direction of research, i.e.\ the study of potentially interesting
effects and phenomena, such as the lensing associated with gravity
waves (Kaiser \& Jaffe 1997) or cosmological applications where
gravitational lensing probes the geometry of the universe (Kaiser
1998; Lombardi \& Bertin 1999, Paper~III; van Waerbeke, Bernardeau, \&
Mellier 1999). Measuring these effects, admittedly at the limits of
current observations, is expected to become feasible with the advent
of the next generation telescopes.

The theory of double lenses has already been subject to some progress.
For example, it has been proven that multiple lenses produce an odd
number of images from a pointlike source (Seitz \& Schneider 1992) and
that a simple theorem on the magnification of images holds (Seitz \&
Schneider 1994), while Crawford et al.\ (1986) have discussed the
probability that a quasar is strongly lensed by two clusters. However,
this still remains a challenging line of research that is so far
largely unexplored; in particular, the problem has not been discussed,
to our knowledge, in the important context of weak lensing analyses.
Attracted by the new developments in deep imaging and especially in
the observation of very distant clusters of galaxies, which already
suggest that a chance alignment of two clusters is not to be
considered an unlikely event (for example, see the case of Cl$0317+15$
noted by Molinari, Buzzoni, \& Chincarini 1996, and of A$1758$
reported by Wang \& Ulmer 1997), we found it natural to give further
thought to such lens configurations.

In this paper, for the context of weak lensing, we generalize the
procedure to extract the astrophysical information contained in the
observed shear field from the case of a single lens to that of a
double lens. We start by showing how the trace and the asymmetric part
of the Jacobian matrix associated with the ray-tracing transformation
can be measured (Sect.~2 and Sect.~3), once the observed ellipticities
of a large number of background galaxies have been properly secured.
We then consider some explicit questions that can be tackled in
configurations of astrophysical interest; in particular, we formulate
and address a ``dark cluster problem,'' corresponding to the situation
where the signature of double lensing is observed at variance with the
apparent absence of a second bright cluster partially aligned along
the line of sight (Sect.~4). We then estimate the size of the effects
involved (Sect.~5), so that we can state that, based on the known
properties of the distribution of bright clusters, a few
configurations are likely to be present in the sky, for which the
small effects characteristic of double lensing may already be within
detection limits. Surprisingly, a discussion of the relevant
contributions to the noise of the shear measurement reveals an
intrinsic limitation of weak lensing analyses. In fact, we find that
the noise contribution associated with the two-galaxy cosmological
correlation function does not decrease with the depth of the
observation. A quantitative evaluation of this subtle effect and of
its impact on mass reconstructions deserves a separate investigation
and will be considered in a future paper. Some numerical tests in
Sect.~6 demonstrate that the analytical framework developed is sound
and bring out the existence of some curious effects (especially on the
criticality condition for a double lens and on the possibility of
pinpointing the location of the invisible cluster in the dark cluster
problem). Then, simulations in the same section show that the redshift
distribution of the source galaxies, which in principle might also
contribute to break the curl-free character of the shear field,
actually produces systematic effects typically two orders of magnitude
smaller than the double lensing effects we are focusing on. In
addition, they suggest that reasonable assumptions on the observed
galaxies lead to noise levels that do not mar the possibility of
detecting the double-lensing effect. Even if, at present, much of what
we have obtained is barely within the limits of direct experimental
confirmation, with the advent of new instrumentation (such as ACS on
HST) and powerful telescopes of the next generation (such as NGST) the
observational impact of analyses of the type provided here is bound to
become significant.


\section{Basic relations}

In this section we briefly recall the lensing equations for double
lenses, in the context where the lensing is produced by intervening
clusters. For the purpose, we mostly follow Schneider, Ehlers, \&
Falco (1992), with some small differences of notation (see also
Paper~III). We then summarize the principles of the statistical
analysis that leads to a local measurement of the gravitational shear.

\subsection{Ray-tracing equation}

Let us consider for the moment a single source at redshift
$z^\mathrm{s}$. Let $D(z_1, z_2)$ be the angular diameter distance
between two aligned objects at redshift $z_1$ and $z_2$, and $D_{ij} =
D\bigl( z^{(i)}, z^{(j)} \bigr)$ (with $i, j \in \{ \textrm{o}, 1, 2,
\textrm{s} \}$ and $0 = z^{(\mathrm{o})} < z^{(1)} < z^{(2)} <
z^{(\mathrm{s})}$) the distances between two of the elements of the
lens configuration, made of observer (o), two deflector planes ($1$
and $2$), and source plane (s). We call $\vec\theta^{(1)}$ (or simply
$\vec\theta$) the apparent position of the source, so that $\vec
x^{(1)} = D_{\textrm{o}1} \vec\theta^{(1)}$ is the corresponding
linear position on the first deflector plane. The light ray is traced
back to angular positions $\vec\theta^{(2)} = \vec x^{(2)} /
D_{\textrm{o}2}$ and $\vec\theta^{(\textrm{s})} = \vec
x^{(\textrm{s})} / D_{\textrm{os}}$ referred to the second deflector
and to the source (see Fig.~\ref{Fig:1}). For simplicity, in the
following we will use the notation $\vec\theta^\mathrm{s}$ for
$\vec\theta^{(\mathrm{s})}$. The ``dynamics'' of the two lenses is
contained in two deflection functions $\vec\alpha^{(1)}$ and
$\vec\alpha^{(2)}$, such that
\begin{align}
  \vec x^{(2)} &= D_{\mathrm{o}2} \vec\theta^{(1)} - D_{12}
  \vec\alpha^{(1)} \bigl( D_{\mathrm{o}1} \vec\theta^{(1)} \bigr) \; ,
  \\
  \vec x^{(\mathrm{s})} &= D_{\mathrm{os}} \vec\theta^{(1)} -
  D_{1\mathrm{s}} \vec\alpha^{(1)} \bigl( D_{\mathrm{o}1}
  \vec\theta^{(1)} \bigr) - D_{2\mathrm{s}} \vec\alpha^{(2)} \bigl(
  D_{\mathrm{o}2} \vec\theta^{(2)} \bigr) \; .
\end{align}
Thus the \textit{ray-tracing equation\/} can be written as
\begin{equation}
  \label{eq:ray-tracing-0}
  \vec\theta^\mathrm{s} = \vec\theta - \vec\beta^{(1)}(\vec\theta) -
  \vec\beta^{(2)} \bigl( \vec\theta - \Delta \vec\beta^{(1)}
  (\vec\theta) \bigr) \; ,
\end{equation}
with ($\ell=1,2$)
\begin{equation}
  \vec\beta^{(\ell)} (\vec\theta) =
  \frac{D_{\ell\mathrm{s}}}{D_\mathrm{os}} \vec\alpha^{(\ell)}
  ( D_{\mathrm{o}\ell} \vec\theta)
\end{equation}
and
\begin{equation}
  \label{eq:Delta}
  \Delta = \frac{D_\mathrm{os} D_{12}}{D_{1\mathrm{s}}
  D_{\mathrm{o}2}} \; .
\end{equation}

For given projected mass distributions of the two lenses
$\Sigma^{(\ell)}$, the distances $D_{ij}$ enter the problem through
$\Delta$ and through the two critical densities, defined as
\begin{equation}
  \label{eq:2}
  \Sigma_\mathrm{c}^{(\ell)} = 
  \begin{cases}
    \infty & \mbox{for $z \le z^{(\ell)}$} \; , \\
    \dfrac{c^2 D_{\mathrm{os}}}{4 \pi G D_{\ell\mathrm{s}}
    D_{\mathrm{o}\ell}} & \mbox{otherwise} \; .
  \end{cases}
\end{equation}
The definition of $\Sigma_\mathrm{c}^{(\ell)}$ for $z \le z^{(\ell)}$
just states that foreground sources are unaffected by the lens.  The
functions $\vec\beta^{(\ell)} (\vec\theta)$ can be expressed by
suitable integrals of the two reduced densities $\kappa^{(\ell)} =
\Sigma^{(\ell)} \bigm/ \Sigma^{(\ell)}_\mathrm{c}$. For cases of
interest, $\Delta$ is smaller than unity. Note that, much like in the
case of a single lens, since distances enter only through the ratios
$D_{1\mathrm{s}} / D_{\mathrm{os}}$ and $D_{2\textrm{s}} /
D_{\mathrm{os}}$, for a family of very distant sources the ray-tracing
equation~\eqref{eq:ray-tracing-0} can be applied without explicit
reference to the distance of each source, and $\Delta$ remains finite.

If the sources are located at different redshifts but not all of them
are at large distance, we can still retain a form similar to
Eq.~\eqref{eq:ray-tracing-0} for the ray-tracing equation provided we
introduce the appropriate \textit{cosmological weight functions\/}:
\begin{equation}
  \label{eq:1}
  w^{(\ell)}(z) =
  \frac{\Sigma^{(\ell)}_\mathrm{c}(z^\mathrm{s})}{\Sigma_\mathrm{c}(z)} \; .
\end{equation}
The functions $w^{(\ell)}(z)$ give the strength of the lens $\ell$ on a
source at redshift $z$ relative to a source at redshift
$z^\mathrm{s}$. [Note that in Paper~III we used a similar definition
of cosmological weight function, but took the ``reference'' redshift
$z^\mathrm{s}$ at infinity. Here we prefer to avoid the limit
$z^\mathrm{s} \rightarrow \infty$.]\@ With this definition,
Eq.~\eqref{eq:ray-tracing-0} can be rewritten as
\begin{equation}
  \label{eq:ray-tracing}
  \vec\theta^\mathrm{s} = \vec\theta - w^{(1)} \vec\beta^{(1)}(\vec\theta) -
  w^{(2)} \vec\beta^{(2)} \bigl( \vec\theta - \Delta \vec\beta^{(1)}
  (\vec\theta) \bigr) \; .
\end{equation}

An important qualitative aspect of the two-lens ray-tracing equation
is the following. In the limit where $\Delta \rightarrow 0$, the
ray-tracing equation adds two contributions,
$\vec\beta^{(1)}(\vec\theta)$ and $\vec\beta^{(2)}(\vec\theta)$, each
deriving from a potential, so that the related Jacobian matrix $A =
\partial \vec\theta^\mathrm{s} / \partial \vec\theta$ is symmetric.
In turn, when $\Delta \neq 0$, the Jacobian matrix is no longer
guaranteed to be symmetric (in fact, the product of two symmetric
matrices is not necessarily symmetric). The limit $\Delta \rightarrow
0$ occurs when the two deflectors are very close to each other.

In the \textit{weak lensing limit\/} the Jacobian matrix is considered
to be close to the identity matrix, $A = \mbox{Id} +
\mathcal{O}(\epsilon)$, with the parameter $\epsilon$ measuring the
strength of the lens. Equation~\eqref{eq:ray-tracing} readily shows
that for two lenses of comparable strength one expects a very small
asymmetry in the Jacobian matrix, with $A_{12} - A_{21} = \mathcal{O}
(\Delta \epsilon^2)$. In the following we will find that, in spite of
the smallness of the effect involved, a weak lensing analysis, carried
out to second order, can provide interesting indications as to the
possibility of detecting, in realistic cases, the presence of a double
lens through its signature of an asymmetry in the Jacobian matrix.
Simulations will show that such indications are further encouraged
when the lenses involved are relatively strong.

\subsection{Local statistical analysis}

Consider a large number $N$ of extended sources all subject to the
\textit{same\/} Jacobian matrix $A$ for the relevant ray-tracing. In
general, as noted above, $A \neq A^\mathrm{T}$. Let $Q_{ij}$ be
the measured quadrupole moments of the individual galaxies. These
observed quantities are related to the source (unlensed) quadrupole
moments $Q^\mathrm{s}_{ij}$ by the equation (see Schneider, Ehlers, \&
Falco 1992; see also Paper~I)
\begin{equation}
  Q^\mathrm{s} = A Q A^\mathrm{T} \; .
\end{equation}
In the limit $N \rightarrow \infty$, the mean value of the source
quadrupoles for an isotropic population of source galaxies should be
proportional to the identity matrix
\begin{equation}
  \langle Q^\mathrm{s} \rangle = M \, \mbox{Id} \; ,
\end{equation}
with $M$ a positive constant. This is the starting point of the
$Q$-method described in Paper~I and allows us to invert the relation
for the mean values
\begin{equation}
  A \langle Q \rangle A^\mathrm{T} = \langle Q^\mathrm{s} \rangle = M
  \, \mbox{Id} \; ,
\end{equation}
into
\begin{equation}
  \label{eq:AA^T}
  A A^\mathrm{T} = M \bigl( \langle Q \rangle \bigr)^{-1} \; .
\end{equation}
Note that the matrix $\bigl( A A^\mathrm{T} \bigr)$ is symmetric. Here
we have taken $\langle Q \rangle$ to be non-singular.

The process of mass reconstruction makes use of the departures from
the identity matrix of the ray tracing matrix as brought out by the
observed quadrupoles. In practice, the true Jacobian matrix $A_0$ is
considered to be unknown. Equation~\eqref{eq:AA^T} shows that if $A_0$
is a solution, any matrix $A = U A_0$ obtained by multiplication by an
orthogonal matrix (so that $U U^\mathrm{T} = \mbox{Id}$) is also a
solution. Thus in the determination of the shear associated with
gravitational lensing the Jacobian matrix is bound to be identified
only up to an orthogonal matrix. It is easy to show that any solution
$A$ for a problem for which $A_0$ is a solution can be written as $A =
U A_0$, with $U$ a suitable orthogonal matrix.

The same set of data $\bigl\{ Q^{(n)} \bigr\}$ can be analyzed by a
standard single lens procedure, leading to the determination of a
symmetric Jacobian matrix $A_\mathrm{s}$, with the property
$A_\mathrm{s} = A_\mathrm{s}^\mathrm{T}$. Then $A_\mathrm{s}$ is to be
related to the true Jacobian matrix $A_0$ by means of an orthogonal
matrix. In fact, for a given $A_0$, there are in general four
different symmetric matrices available (corresponding to matrices
that differ by the sign of their eigenvalues). If we restrict the
attention to transformations with positive determinant (the $\det U =
\pm 1$ ambiguity reflects the well-known $g \mapsto 1/g^*$ invariance;
see Schneider \& Seitz 1995), we can write
\begin{equation}
  A_\mathrm{s} = 
  \begin{pmatrix}
    \cos \tau  & \sin \tau \\
    -\sin \tau & \cos \tau
  \end{pmatrix}
  A_0 \; ,
\end{equation}
with
\begin{equation}
  \label{eq:tantau}
  \tan \tau = - \frac{\varepsilon_{k k'} A_{0 k k'}}{ A_{0 m m} } \; .
\end{equation}
Here $\varepsilon_{ij}$ is the totally antisymmetric tensor of rank
2 and $A_{0 k k} = \func{Tr}(A_0)$ is the trace of $A_0$ (we use the
summation convention on repeated indices). Note that the quantity
$\varepsilon_{k k'} A_{0 k k'} = A_{012} - A_{021}$ is a measure of
the asymmetry of $A_0$.

The symmetric matrix can thus be given in explicit form:
\begin{equation}
  \label{eq:+-f}
  A_{\mathrm{s}ij} = \pm \frac{1}{f} \bigl( A_{0ij} A_{0mm} +
  \varepsilon_{j j'} A_{0 i j'} \varepsilon_{k k'} A_{0 k k'}
  \bigr) \; ,
\end{equation}
with
\begin{equation}
  f^2 = (A_{0 m m})^2 + (\varepsilon_{k k'} A_{0 k k'})^2 \; .
\end{equation}

In what has been described so far, $A$ is considered to be the same
for all the source galaxies. Therefore at this stage we have addressed
only the problem of a {\it local} analysis, applicable to source
galaxies very close to each other, in a small patch of the sky, and at
a similar redshift. In conclusion, we have so far shown that a local
measurement cannot lead to discovering a double lens, because there is
always a way to interpret the data by means of a symmetric ray-tracing
Jacobian matrix.


\section{Non-local analysis in the weak lensing limit}

Let us first consider only sources located at a given redshift $z$.
In the weak lensing limit, the Jacobian matrix for such sources
is of the form $A_0 = \mbox{Id} + \mathcal{O}(\epsilon)$, and the
asymmetry is small, $\varepsilon_{k k'} A_{0 k k'} =
\mathcal{O}(\epsilon^2)$. Thus we can write the related symmetric
matrix (the ``effective'' Jacobian matrix) as
\begin{equation}
  \label{eq:B17}
  A_\mathrm{s} = 
  \begin{pmatrix}
    \sigma + \gamma_1 & \gamma_2 \\
    \gamma_2 & \sigma - \gamma_1
  \end{pmatrix}
  \simeq
  \begin{pmatrix}
    A_{011} & A_{012} + \tau \\
    A_{021} - \tau & A_{022}
  \end{pmatrix}
\end{equation}
with (see Eq.~\eqref{eq:tantau})
\begin{equation}
  \label{eq:B18}
  \tau \simeq - \frac{1}{2} ( A_{012} - A_{021}) =
  \mathcal{O}(\epsilon^2) \; .
\end{equation}
Therefore the elements of the symmetric matrix (appearing in the mass
reconstruction analysis) can be expressed in terms of those of the
true Jacobian matrix, as
\begin{align}
  \label{eq:B19}
  \sigma &= \frac{1}{2} \func{Tr} (A_0) \; , \\
  \gamma_1 &= \frac{1}{2} (A_{011} - A_{022}) \; , \\
  \label{eq:B22}
  \gamma_2 &= \frac{1}{2} (A_{012} + A_{021}) \; .
\end{align}
From inspection of Eqs.~\eqref{eq:B18} and \eqref{eq:B19} and from the
definition of the \textit{true\/} Jacobian matrix, we find that the
\textit{true\/} source positions are related to $\sigma$ and to $\tau$
by
\begin{align}
  \nabla \cdot \vec\theta^\mathrm{s} &= \func{Tr}(A_0) = 2 \sigma \; ,
  \\
  \nabla \wedge \vec\theta^\mathrm{s} &= - (A_{012} - A_{021}) = 2
  \tau \; .
\end{align}
Thus, if we refer to the vector field $\vec u(\vec\theta)$ commonly
used in weak lensing analyses, we find
\begin{equation}
  \vec u = - 
  \begin{pmatrix}
    \gamma_{1,1} + \gamma_{2,2} \\
    \gamma_{2,1} - \gamma_{1,2}
  \end{pmatrix}
  = - \frac{1}{2} \nabla^2 \vec\theta^\mathrm{s} \; .
\end{equation}
If we now combine the definition of $\vec u(\vec\theta)$ with the
above relations for $\sigma$ and $\tau$ we get
\begin{align}
  \label{eq:B23}
  \nabla \cdot \vec u &= - \nabla^2 \sigma \; , \\
  \label{eq:B24}
  \nabla \wedge \vec u &= - \nabla^2 \tau \; .
\end{align}
The last two equations replace the well-known relations applicable to
single lens analyses, i.e.\ $\vec u = - \nabla \sigma$ and $\nabla
\wedge \vec u = 0$.

Suppose now that the source galaxies follow a redshift distribution
$p(z)$. At each redshift, we can apply
Eqs.~(\ref{eq:B17}--\ref{eq:B24}) provided that all lensing
quantities, such as $A_0$, $\sigma$, $\gamma$, and $\tau$, are
calculated for each source at the correct redshift. In the weak
lensing limit, the shear $\gamma$ is estimated from the ellipticities
of galaxies that are observed in a particular region of the sky.
Since source galaxies have different redshifts, in reality a mean
value of the shear is measured:
\begin{equation}
  \label{eq:3}
  \langle \gamma \rangle_z = \int_0^\infty \gamma(z) p(z) \, \diff z
  \; ,
\end{equation}
where $\gamma(z)$ is the redshift-dependent shear. In the weak lensing
limit, all quantities depend linearly on the shear, and thus
Eqs.~\eqref{eq:B23} and \eqref{eq:B24} can be written as
\begin{align}
  \label{eq:B23-z}
  \nabla \cdot \langle \vec u \rangle_z &= - \nabla^2 \langle \sigma
  \rangle_z \; , \\
  \label{eq:B24-z}
  \nabla \wedge \langle \vec u \rangle_z &= - \nabla^2 \langle \tau
  \rangle_z \; .
\end{align}

Unfortunately, these relations cannot be used to infer directly, from
a given set of data, the amount of asymmetry that would be a
characteristic signature of multiple lensing. The reason is that for
this purpose the data should be able to identify the vector field
$\vec u(\vec\theta)$ up to $\mathcal{O}(\epsilon^2)$, because we know
that $\tau = \mathcal{O}(\epsilon^2)$. In practice, it is well-known
that standard weak lensing analyses lead to the determination of $\vec
u(\vec\theta)$ only up to $\mathcal{O}(\epsilon)$. In fact,
observations lead to the determination of the \textit{reduced shear\/}
$g(\vec\theta) = \gamma(\vec\theta) / \sigma(\vec\theta)$; the
identification of $g$ with $\gamma$ is correct only to
$\mathcal{O}(\epsilon)$. Thus the above analysis is not yet ready for
practical applications. For this reason we need to extend the
discussion so as to include the \textit{second order\/} terms in the
weak lensing expansion. In principle, the discussion can be carried
out by retaining the redshift dependence of sources. In practice, such
approach would lead us far beyond the original purpose of this paper.
For this reason we will suppose, in the following, that all sources
are located at redshift $z^\mathrm{s}$ (in Sect.~5 we will consider
again the spread of sources in redshift for the single lens case).

For a population of source galaxies located at a single redshift (even
in the case of strong lensing; but in this section, we recall, we are
still within the weak lensing expansion) an ``observable'' field is
(see Kaiser 1995)
\begin{equation}
  \label{eq:u_tilde}
  \vec{\tilde u} = \frac{1}{1 - |g|^2} 
  \begin{pmatrix}
    1 + g_1 & g_2 \\
    g_2 & 1 - g_1
  \end{pmatrix}
  \begin{pmatrix}
    g_{1,1} + g_{2,2} \\
    g_{2,1} - g_{1,2}
  \end{pmatrix} \; ;
\end{equation}
for a strong single lens, such field has the important property that
$\vec{\tilde u} = \nabla \bigl[ \ln \bigl| 1 - \kappa(\vec\theta)
\bigr| \bigr]$, with $\kappa(\vec\theta)$ the dimensionless projected
density that one aims at reconstructing. Curiously, \textit{to second
  order\/} in the weak lensing expansion (see Appendix~A), it is
possible to show that, for multiple weak lenses of comparable
strength,
\begin{align}
  \label{eq:B26}
  \nabla \cdot \vec{\tilde u} &= \nabla^2 ( \ln \sigma ) \; , \\
  \label{eq:B27}
  \nabla \wedge \vec{\tilde u} &= \nabla^2 \tau \; .
\end{align}
In other words, \textit{we have derived a set of equations for which
  the contribution of a small asymmetry present in the true Jacobian
  matrix is retained consistently\/}; the structure is similar to that
of the set for $\vec u(\vec\theta)$ field presented in
Eqs.~\eqref{eq:B23} and \eqref{eq:B24}. \textit{The new set of
  Eqs.~\eqref{eq:B26} and \eqref{eq:B27} is the basis that allows us
  to generalize the procedure to extract the available information,
  contained in the observed shear field, from the case of a single
  lens to that of a multiple lens and thus to investigate
  quantitatively the characteristics of the coupling of two or more
  deflectors located along the same line of sight}. The symmetric
limit, which we may call in this context the single lens limit, is
easily recognized.

The results of this section (Eqs.~\eqref{eq:B26} and \eqref{eq:B27})
are not generalized easily to the case of sources distributed
according to a $p(z)$. However, the simulations described in Sect.~6.2
will basically support a description analogous to that given by
Eqs. \eqref{eq:B23-z} and \eqref{eq:B24-z}.

The conclusions of this section, that the data contained in the
\textit{shear map\/} can be used to detect an asymmetry in the true
Jacobian matrix $A_0$, and of the previous Sect.~2.2, that such a
detection is impossible if based on a local measurement only, do not
depend on the number of deflectors involved in a multiple lens. In the
following, we will specialize our conclusions to the study of double
lenses, for which the ray-tracing equation can be handled in a
straightforward manner.


\section{Double lenses in the astrophysical context}

In this section we formulate some questions that may be interesting
from the astrophysical point of view. The expected size of some
effects and the possibility of an actual measurement will be addressed
in separate sections at the end of the paper.

\subsection{Where to look for the signature of double lensing?}

The prime signature of double lensing would be the detection of a
significant asymmetry $\tau$ (see Eq.~\eqref{eq:B27}). Because of
statistical errors, the measured vector field $\vec{\tilde u}$ is
bound to be associated with a non-vanishing curl. Therefore, a
positive detection of asymmetry can be claimed only if the expected
statistical error on $\tau$ is smaller than the true value $\tau_0$
characterizing the double lens.

A quantitative analysis of this condition will be provided in Sect.~5
below. Here we only note that, in general, for two clusters of
comparable strength, at different distances, with offset centers with
respect to the line of sight, the regions on the sky where the
signal-to-noise ratio for $\tau$ should be largest are those, on
either side, just off the line connecting the cluster centers (see
Fig.~\ref{Fig:density2} and Fig.~\ref{Fig:dens_plot})

As we have seen in Sect.~2, the asymmetry is expected to be weak
($\mathcal{O}(\Delta \epsilon^2)$ for two clusters of comparable
strength $\epsilon$). The optimal conditions for detecting
double lensing are then:
\begin{itemize}
\item The geometric parameter $\Delta$ should be close to unity, i.e.,
  for a given distribution of sources, the distance between the two
  clusters should not be much smaller than the distance from the
  observer to the near cluster. This condition is reasonably satisfied
  if, for example, the redshifts of the two clusters are in the
  following relation, $z^{(2)} \simeq 4 z^{(1)}$.
\item The two clusters should be not too weak. It is preferable to
  consider cases where the dimensionless densities, $\kappa^{(1)}$ and
  $\kappa^{(2)}$, are of order unity.
\item The cluster centers should be offset, but there should be a
  region of significant overlap between the two clusters in the sky.
\end{itemize}

\subsection{Probability of cluster alignments}

Given the above criteria, we now estimate the probability that a
double lens with desired properties be observed. For the purpose, we refer
to $H_0 = 65 \mbox{ km s}^{-1} \mbox{ Mpc}^{-1}$ for the Hubble
constant, within a Friedmann-Lema\^\i tre cosmological model
characterized by $\Omega = 0.3$ and $\Omega_\Lambda = 0.7$.

First, we suppose that a nearby cluster is observed at redshift
$z^{(1)} \approx 0.1$ (the values of the various quantities used here
are similar to the values used in the simulations to be described in
Sect.~\ref{sect:simulations}) and ask what is the probability of
finding a cluster at redshift $z^{(2)} \approx 4 z^{(1)}$ well aligned
with the first cluster; in considering this question, we further
require that the second cluster be sufficiently massive, with mass
greater than $5 \times 10^{14} \mbox{ M}_\odot$. For the purpose, we
assume that the angular distance between the centers of the two
clusters be between $2$ and $4\mbox{ arcmin}$ and that the redshift of
the second cluster be in the range $0.3 \leq z^{(2)} \leq 0.5$. A
simple calculation then shows that the center of the second cluster
must be inside a comoving volume with size $V \approx 10\, 360 \mbox{
  Mpc}^3$. Therefore, based on the cluster density found by Borgani et
al. (1999; see also Girardi et al.\ 1998), the expected probability of
finding a ``good'' double lens turns out to be $\approx 0.3 \%$. If we
consider less strict requirements, with an angular distance between
the two cluster centers in the range $1$--$5\mbox{ arcmin}$ and a
redshift range $0.2$--$0.6$ for the second cluster, the estimated
probability increases up to $\approx 1 \%$.

These estimates should be combined with the number of massive clusters
at small redshift (for the object observed at $z^{(1)}$) that should
be available over the whole sky. The comoving volume for a shell with
redshift between $0.05$ and $0.15$ is $\approx 1.34 \times 10^9 \mbox{
  Mpc}^3$; correspondingly, the total number of clusters with mass
above $5 \times 10^{14} \mbox{M}_\odot$ in such a volume is about
$350$. As a result, we expect from $1$ to $3$ cases of double lenses
with detectable effects and configurations similar to the one
considered in our simulations. From the way we have approached the
problem, it is clear that the estimate just obtained provides only a
\textit{lower limit} with respect to the number of good cases expected
for the purposes of the present paper.

\subsection{Mass reconstruction for double lenses}

Suppose that a set of data leads to the determination of $\vec{\tilde
  u}$ and that, based on suitable boundary conditions (e.g.,
$\tau(\vec\theta) \rightarrow 0$, $\sigma(\vec\theta) \rightarrow 1$
for $\| \vec\theta \| \rightarrow \infty$), we integrate
Eqs.~\eqref{eq:B26} and \eqref{eq:B27} and get the functions
$\sigma(\vec\theta)$ and $\tau(\vec\theta)$. For a single lens problem
the lensing analysis, at this stage, would be complete, since we would
have the dimensionless projected mass distribution $\kappa(\vec\theta)
= 1 - \sigma(\vec\theta)$ associated with the deflector; knowledge of
the critical density $\Sigma_\mathrm{c}$ would lead to the projected
mass distribution $\Sigma(\vec\theta)$. In the case of a double lens,
what would be the astrophysically interesting quantities that could be
derived?

For the following discussion, the ray-tracing equation can
be re-cast in a more transparent form in the case $z = z^{(s)}$ (so
that $w^{(1)} = w^{(2)} = 1$). If we introduce the two
\textit{curl-free lensing maps\/} (we recall that $\nabla \wedge
\vec\beta^{(\ell)} = 0$)
\begin{align}
  \label{eq:zeta1}
  \vec\zeta^{(1)} (\vec\theta) &= \vec\theta - \Delta \vec\beta^{(1)}
  (\vec\theta) \; , \\
  \label{eq:zeta2}
  \vec\zeta^{(2)} (\vec\theta) &= \vec\theta - \Delta \vec\beta^{(2)}
  (\vec\theta) \; ,
\end{align}
and the \textit{effective ray-tracing map\/}
\begin{equation}
  \label{eq:eff-ray-tracing}
  \vec\zeta^\mathrm{s} (\vec\theta) = (1 - \Delta) \vec\theta + \Delta
  \vec\theta^\mathrm{s}(\vec\theta) \; ,
\end{equation}
we see that Eq.~\eqref{eq:ray-tracing} can be written as
\begin{equation}
  \label{eq:ray-tracing-2}
  \vec\zeta^\mathrm{s}(\vec\theta) = \vec\zeta^{(2)} \bigl
  ( \vec\zeta^{(1)} (\vec\theta) \bigr) \; .
\end{equation}
In other words, ray-tracing for a double lens can be seen as the
composition of two curl-free functions.

The double lens problem in its \textit{direct\/} formulation involves
two (dimensionless) ray-tracing functions
($\vec\beta^{(1)}(\vec\theta)$ and $\vec\beta^{(2)}(\vec\theta)$), two
critical densities ($\Sigma_\mathrm{c}^{(1)}$ and
$\Sigma_\mathrm{c}^{(2)}$), and one additional geometric parameter
$\Delta$. Any inverse problem would be under-determined, if we start
from the two dimensionless functions $\sigma(\vec\theta)$ and
$\tau(\vec\theta)$ alone. Even if the geometry is taken to be known
(for example, when the distances to the two clusters acting as
deflectors are known), it is not possible from the pair $(\sigma,
\tau)$ to ``reconstruct the two lenses,'' i.e.\ to get the ray-tracing
functions $\vec\beta^{(1)}$ and $\vec\beta^{(2)}$ separately. This
statement is easy to understand, especially if we refer to the
curl-free lensing maps $\vec\zeta^{(1)}$ and $\vec\zeta^{(2)}$ of
Sect.~2.1. In fact, a simple scaling invariance $\vec\zeta^{(1)}
\mapsto k \vec\zeta^{(1)}$, $\vec\zeta^{(2)} \mapsto k^{-1}
\vec\zeta^{(2)}$ shows that, unless we are able to provide suitable
boundary conditions on the functions $\vec\zeta^{(\ell)}(\vec\theta)$
(e.g., $\vec \zeta^{(2)}(\vec\theta) \rightarrow 1$ for large $\|
\vec\theta \|$), there is no way to disentangle from the shear data
alone the contributions of the two lenses. Obviously, this does not
mean that the lensing analysis is useless; it is only a reminder that
in this more complex situation an unambiguous mass reconstruction
based on weak lensing would require additional input from other probes
of mass distributions (e.g., X-ray data).

In order to clarify this point further, it may be instructive to
consider the following simple examples, all characterized by $\tau =
0$.

\subsubsection{Aligned, centrally symmetric lenses}

Suppose that
\begin{equation}
  \vec\zeta^\mathrm{s}(\vec\theta) = \frac{\vec\theta}{\| \vec\theta \|}
  \zeta^\mathrm{s} \bigl( \| \vec\theta \| \bigr) \; ,
\end{equation}
with $\zeta^\mathrm{s}( \theta )$ a real continuous function such that
$\zeta^\mathrm{s}(0) = 0$. Then it is natural to choose
$\vec\zeta^{(1)}$ and $\vec\zeta^{(2)}$ to be radial, i.e.
\begin{align}
  \vec\zeta^{(1)}(\vec\theta) &= \frac{\vec\theta}{\| \vec\theta \|}
  \zeta^{(1)} \bigl( \| \vec\theta \| \bigr) \; , \\
  \vec\zeta^{(2)}(\vec\theta) &= \frac{\vec\theta}{\| \vec\theta \|}
  \zeta^{(2)} \bigl( \| \vec\theta \| \bigr) \; ,
\end{align}
where $\zeta^{(1)}$ and $\zeta^{(2)}$ are real functions. Supposing
that $\zeta^{(1)}(\theta) \ge 0$ for all $\theta$, we can rewrite
Eq.~\eqref{eq:ray-tracing-2} as
\begin{equation}
  \zeta^{(2)} \bigl( \zeta^{(1)}( \theta ) \bigr) =
  \zeta^\mathrm s (\theta) \; .
\end{equation}
Thus reduced to a one-dimensional problem, it is clear that this
equation in general admits an infinite number of solutions.

\subsubsection{The general case of double lenses without asymmetry
  ($\tau = 0$)}

Consider an effective ray-tracing map (see
Eq.~\eqref{eq:eff-ray-tracing}) such that $\nabla \wedge
\vec\zeta^\mathrm{s} = \nabla \wedge \vec\theta^\mathrm{s} = 0$. In
this case we can solve Eq.~\eqref{eq:ray-tracing-2} by choosing an
invertible curl-free map $\vec\zeta^{(2)}$ and by writing
\begin{equation}
  \vec\zeta^{(1)} (\vec\theta) = \bigl( \vec\zeta^{(2)} \bigr)^{-1}
  \bigl( \vec\zeta^\mathrm{s} (\vec\theta) \bigr) \; .
\end{equation}
We now require that $\vec\zeta^{(1)}$ be curl-free. From the above
equation, this happens if and only if the product of the Jacobian matrix
$A^\mathrm{(-2)}$, associated with $\bigl( \vec\zeta^{(2)}
\bigr)^{-1}$, and $A^\mathrm{s}$, associated with
$\vec\zeta^\mathrm{s}$, is a symmetric matrix, i.e.
\begin{equation}
  \label{eq:B35}
  A^{(-2)} A^\mathrm{s} = \bigl( A^{(-2)} A^\mathrm{s} \bigr)^T =
  A^\mathrm{s} A^{(-2)} \; .
\end{equation}
From linear algebra we know that this is equivalent to saying that
$A^{(-2)}$ and $A^\mathrm{s}$ have the same eigenvectors. Since
$\vec\zeta^{(2)}$ is curl-free, let us introduce the potential
$\phi^{(2)}$ so that $\vec\zeta^{(2)} = \nabla \phi^{(2)}$ and
$A^{(2)}_{ij} = \phi_{,ij}^{(2)}$. If $\vec\xi = (\xi_1, \xi_2)$ is a
\textit{local\/} system of coordinates where $A^\mathrm{s}$ is
diagonal, then Eq.~\eqref{eq:B35} is satisfied if $A^{(2)}$ is
diagonal in the same system of coordinates, i.e.\ if $\partial^2
\phi^{(2)} \bigm/ \partial \xi_1 \partial \xi_2 = 0$. This simply
states that $\phi^{(2)}$ is \textit{separable\/} in $\xi_1$ and
$\xi_2$, i.e.\ that we can write $\phi^{(2)}$ as the sum of an
arbitrary function of $\xi_1$ and of an arbitrary function of $\xi_2$.
This again demonstrates the freedom at our disposal in solving
Eq.~\eqref{eq:ray-tracing-2}. In addition, one can now better
appreciate why, even when natural boundary conditions are specified,
in general there is no guarantee that the solution be determined
uniquely; in fact, several solutions are expected when the coordinate
system $(\xi_1, \xi_2)$ is associated with \textit{poles\/} (such as
the point $r = 0$ for polar coordinates).

\subsubsection{Two lenses with no net lensing?}

One curious case that may be imagined is the possibility of combining
two lenses with no net effect, so that $\vec\zeta^\mathrm{s}
(\vec\theta) = \vec\theta$ (see Eq.~\eqref{eq:eff-ray-tracing}). In
principle, one may argue that from any invertible curl-free map
$\vec\zeta^{(1)}(\vec\theta)$ one can take a second lens characterized
by $\vec\zeta^{(2)} (\vec\theta) = \bigl(\vec\zeta^{(1)} (\vec\theta)
\bigr)^{-1}$. This idea would obviously generate an infinite number of
solutions since the starting function $\vec\zeta^{(1)}$ is at our
disposal. However, if we require that the density distributions
associated with the two lenses be both \textit{positive definite},
i.e. that $\nabla \cdot \vec\zeta^{(i)} \leq 2$, it can be shown that
there is no way for two lenses to exactly compensate for each other:
in other words, the only admissible solution is the trivial
$\vec\zeta^{(1)}(\vec\theta) = \vec\zeta^{(2)} (\vec\theta) =
\vec\theta$. This conclusion is consistent with the theorem that
states that any combination of gravitational lenses is bound to
produce a net magnification ($\mu > 1$; see Seitz \& Schneider 1992).

\subsection{The dark cluster problem}

A different problem based on double lenses can be formulated in the
following way. Suppose that one observes a lensing cluster and that in
the process of producing the mass reconstruction one finds evidence
that the vector field $\vec{\tilde u}$ is not curl-free, well above
the expected errors. Clearly, in such a situation, one possibility is
that a separate mass concentration, which we may call a \textit{dark
  cluster}, is responsible for the effect. We may now consider the
case when the mass distribution of the visible cluster is well
constrained by diagnostics independent of lensing (e.g., by X-ray
data). Under these circumstances, what can we tell about the dark
cluster properties from the observed lensing effects? In particular,
is it possible to derive the location and the mass distribution of the
invoked dark cluster?

If the mass distribution of the luminous cluster is taken to be known,
one has either the function $\vec\zeta^{(1)}_\Delta$ or the function
$\vec\zeta^{(2)}_\Delta$, depending on whether the dark cluster is
near or far; in either case, the deflection associated with the dark
cluster $\vec\beta^{(d)} = (\vec\theta - \vec\zeta^{(d)}) / \Delta$
depends implicitly (see
Eqs.~\eqref{eq:zeta1}--\eqref{eq:eff-ray-tracing}) on the geometric
parameter $\Delta$, which is unknown. One way to obtain the value of
$\Delta$ is to impose that the lensing map associated with the dark
cluster is curl-free, i.e.\ that either
\begin{equation}
  \label{eq:findz1}
  \nabla \wedge \left[ \vec\zeta^\mathrm{s}_\Delta \bigl
  ( \bigl( \vec\zeta^{(1)}_\Delta \bigr)^{-1} (\vec\theta) \bigr)
  \right] = 0 \; ,
\end{equation}
or that
\begin{equation}
  \label{eq:findz2}
  \nabla \wedge \left[ \vec\zeta^{(2)}_\Delta \bigl
  ( \bigl(\vec\zeta^\mathrm{s}_\Delta \bigr)^{-1} (\vec\theta) \bigr)
  \right] = 0 \; ,
\end{equation}
depending on whether we guess the dark cluster to be near or far. In
Sect.~6 below, by means of a simulated case, we will show how $\Delta$
can be determined with reasonable accuracy, by minimizing the square
of the left side of the above equations. In other words, under the
above circumstances \textit{we will demonstrate that a weak lensing
  analysis allows us to pinpoint the location of the invisible cluster
  and to reconstruct its mass distribution}.

In closing this section, we may note that the problem of a
\textit{single\/} dark cluster, i.e.\ the case of a matter
concentration detected by lensing effects without a visible
counterpart for the lens, is less constrained and, in this respect,
less interesting than the case discussed above in the sense that the
usual mass reconstruction would only lead to the dimensionless
projected density $\kappa$, with no hope to derive the distance to the
dark cluster that is invoked and the actual scale of the mass
involved.


\section{Size of the double lensing effect}

In this section we calculate the expected order of magnitude for the
measurable effects associated with double lensing. For simplicity, the
following section discusses the noise properties of quantities related
to the vector $\vec u$. The results obtained are the leading order
estimates for the desired quantities related to $\vec{\tilde u}$. The
following subsection deals with a population of source galaxies
located at a single redshift. Effects related to a spread in redshift
will be estimated in Sect.~5.3.

\subsection{Expected variance of $\nabla \wedge \vec u$ for a single
  lens}
\label{sec:expect-vari-nabla}

Consider a single lens, characterized by true Jacobian matrix $A_0$,
so that the shear field $\vec u_0$ has vanishing curl, i.e. $\tau_0 =
0$. Because of the finite number of source galaxies used and of the
smoothing introduced in the reconstruction process, we expect that the
measured $\vec u$ in general differs from the true field $\vec u_0$. 

In order to calculate the expectation value and the variance of
$\nabla \wedge \vec u$, we may use a technique similar to that
described in Paper~II. We first recall that in the reconstruction
process Eq.~\eqref{eq:AA^T} is used together with a (positive) weight
function $W (\vec\theta, \vec\theta')$. Such weight function, defined
so that $W(\vec\theta, \vec\theta')$ is significantly different from
zero only for $\vec\theta'$ close to $\vec\theta$, enters in the local
averages of quadrupoles (or ellipticities). In particular, if we
resort to the so-called Q-method associated with Eq.~\eqref{eq:AA^T}
(see Paper~I), the local average quadrupole $\langle Q \rangle
(\vec\theta)$ is calculated from the relation
\begin{equation} 
  \label{eq:<Q>}
  \langle Q \rangle (\vec\theta) = \frac{\sum_n W \bigl(\vec\theta,
  \vec\theta^{(n)} \bigr) Q^{(n)}}{\sum_n W \bigl(\vec\theta,
  \vec\theta^{(n)} \bigr)} \; .
\end{equation}
Here the superscripts $(n)$ label the galaxies affected by lensing. In
Paper~II we have calculated expectation values and variances of
several quantities, under the hypothesis that the weight function is
invariant upon translation, so that $W(\vec\theta, \vec\theta') =
W(\vec\theta - \vec\theta')$, and that it is {\it normalized\/}, i.e.
\begin{equation}
  \label{eq:11}
  \rho \int W(\vec\theta) \, \diff^2\theta = 1 \; . 
\end{equation}
Here $\rho$ is the density of galaxies (taken to be constant over the
field). In particular, we have shown that the expectation values of
the relevant quantities (in particular, the measured shear, the field
$\vec u$, and the reconstructed mass density $\kappa$) are equal to
the true quantities {\it smoothed\/} by the weight function, e.g.
\begin{equation}
  \langle \vec u \rangle(\vec\theta) = \rho \int W(\vec\theta -
  \vec\theta') \vec u_0(\vec\theta') \, \diff^2 \theta' \; .
  \label{eq:<u>}
\end{equation}
From this expression it can be easily shown that the expected
value for the measured curl of $\vec u$ vanishes. In fact,
\begin{equation}
  \begin{split}
    \langle \nabla \wedge \vec u \rangle(\vec\theta) &= \rho
    \nabla_{\vec\theta} \wedge \int W(\vec\theta - \vec\theta') \vec
    u_0(\vec\theta') \, \diff^2 \theta' \\ 
    &= \rho \int W(\vec\theta') \nabla_{\vec\theta} \wedge \vec
    u_0(\vec\theta - \vec\theta') \, \diff^2 \theta' = 0 \; .    
  \end{split}
\end{equation}
Here we have used the symmetry property of convolutions. Thus, for a
single lens, we expect to measure (in the average) a field $\vec u$
with vanishing curl. Note that, for a double lens, a similar
calculation would show that
\begin{equation}
  \label{eq:B42}
  \langle \tau \rangle(\vec\theta) = \rho \int W(\vec\theta -
  \vec\theta') \tau_0(\vec\theta') \, \diff^2 \theta' \; .
\end{equation} 

Let us now turn to covariances. Using the same notation as in
Paper~III (see also Lombardi~2000), we call $c$ the covariance of the
ellipticity distribution for the population of source galaxies
($\langle \epsilon_i^\mathrm{s} \epsilon_j^\mathrm{s} \rangle = c
\delta_{ij}$).  There we have shown that the covariance of the shear
$\gamma$ is
\begin{equation}
  \label{eq:4}
  \begin{split}
    \Cov_{ij}(\gamma; \vec\theta, \vec\theta') &\equiv \bigl\langle \bigl
    ( \gamma_i(\vec\theta) - \langle \gamma_i \rangle(\vec\theta) \bigr)
    \bigl ( \gamma_j(\vec\theta') - \langle \gamma_j \rangle(\vec\theta')
    \bigr) \bigr\rangle \\
    &= c \rho \delta_{ij} \! \int \! W(\vec\theta
    - \vec\theta'') W(\vec\theta' - \vec\theta'') \diff^2 \theta'' \; .
  \end{split}
\end{equation}
Note that the covariance of $\gamma$, which we may write in short as
$\Cov(\gamma)$, depends only on $\vec\theta - \vec\theta'$. Then the
covariance of $\vec u$ is related to the covariance of $\gamma$ in a
simple way:
\begin{equation}
  \Cov (u) = - \nabla^2 \Cov(\gamma) \; .
\end{equation}
Similarly, it can be shown that
\begin{equation}
  \Cov (\nabla \wedge \vec u) = - \nabla^2 \Cov(u) = \nabla^2 \nabla^2
  \Cov(\gamma) \; .
\end{equation}

The last expression is the basis of the following discussion. In fact,
we may argue that a lens can be considered to be double only if the
measured value of $\nabla \wedge \vec u$ is significantly larger than
the expected error on this quantity. If we refer to
Eqs.~\eqref{eq:B18}, \eqref{eq:B19} as definitions of $\sigma$ and
$\tau$, we thus find (under optimal conditions defined in Paper~II)
\begin{equation}
  \label{eq:B46}
  \Cov(\sigma) = \Cov(\tau) = \Cov(\gamma) \; .
\end{equation}
A more useful expression can be found in the simple case of a Gaussian
weight function:
\begin{equation}
  W(\vec\theta, \vec\theta') = \frac{1}{2 \pi \rho \sigma_W^2} \exp
  \left( - \frac{\| \vec\theta - \vec\theta' \|^2}{2 \sigma^2_W} \right)
  \; .
\end{equation}
In this case we have
\begin{align}
  \label{eq:B48}
  \Cov_{ij}(\gamma; \vec\delta) =& \frac{c \delta_{ij}}{4 \pi \rho
  \sigma_W^2} \exp \left( - \frac{\| \vec\delta \|^2}{2 \sigma^2_W}
  \right) \; , \\
  \Cov(\nabla \wedge \vec u; \vec\delta) =& \frac{c \left[ \bigl( 8
  \sigma_W^2 - \| \vec\delta \|^2 \bigr)^2 - 32 \sigma_W^4
  \right]}{64 \pi \rho \sigma_W^{10}} \times {} \nonumber \\
  & {} \times \exp \left( - \frac{\| \vec\delta \|^2}{ 2 \sigma^2_W}
  \right) \; ,
\end{align}
where we have called $\vec\delta \equiv \vec\theta - \vec\theta'$.
Note that $\nabla \wedge \vec u$ shows anti-correlation for $\| \vec
\delta \|$ in the range $\bigl(2 \sqrt{2 - \sqrt{2}}, 2 \sqrt{2 +
  \sqrt{2}} \bigr) \sigma_W$. Similar noise properties are associated
with the quantity $\nabla \cdot \vec{u}$.

\subsection{Condition for the detection of double lensing}

Now we are finally able to set a quantitative condition for the
detection of double lensing. The requirement that the error on $\tau$
(which can be estimated from Eqs.~\eqref{eq:B46} and \eqref{eq:B48})
be smaller than the expected value of $\tau$, within factors of order
unity, can be written as:
\begin{equation}
  \label{eq:B50}
  \Delta \kappa^{(1)} \kappa^{(2)} \gtrsim \sqrt{ \frac{c}{\rho
  \sigma_W^2}} \; .
\end{equation}
Note that $\rho \sigma_W^2$ represents the number of galaxies
effectively used in the average of Eq.~\eqref{eq:<Q>}. If we refer to
a case with $\kappa^{(1)} \simeq \kappa^{(2)} \simeq \Delta \simeq
1/2$ and to the currently ``realistic'' density $\rho \simeq 100
\mbox{ gal arcmin}^{-2}$, with $c \simeq 0.03$, we thus find that a
significant signal requires the use of a Gaussian weight function with
smoothing size $\sigma_W$ not smaller that $10 \arcsec$. On the other
hand, the interesting effects are those associated with the structure
of the field $\tau(\vec\theta)$, which is associated with a length
scale at best comparable with that of the clusters under
investigation.  Therefore, we may argue that the most promising way to
identify the effects of double lensing is to work with the smallest
smoothing size $\sigma_W$ compatible with condition \eqref{eq:B50}.
These arguments and the applicability of the condition defined by
Eq.~\eqref{eq:B50} are clarified and further supported by the
simulations that will be presented in Sect.~6.

\subsection{Effects related to the spread in the distances of the
  source galaxies}

So far, in this section, we have referred to a sheet of source
galaxies at a given distance.  In real situations where the source
galaxies are distributed at different redshifts, we expect three
separate effects that can have a significant impact on our analysis:

\begin{itemize}
\item An extra source of noise, related to the distribution in
  redshift of the individual sources, is added to our data. As a
  result, the measured shear will have a larger covariance than the
  one calculated in Eq.~\eqref{eq:4}, and thus a larger covariance is
  also expected for $\nabla \wedge \vec u$.
\item An additional effect is induced by the extra-clumping of the
  sources described by the cosmological two-point correlation function
  (see Peebles 1980).
\item Outside the limit of very weak lensing, there is actually no
  guarantee that $\nabla \wedge \vec{\tilde u} = 0$ \textit{even in
    the standard case of a single lens}. This happens because the
  observable is, in this case, $\langle g \rangle_z$, which is not
  simply related to $\kappa_0$ and $\gamma_0$.
\end{itemize}
The first two items contribute to the estimate of the relevant
covariances while the third item adds a bias (to be discussed in
Sect.~5.3.2).

\subsubsection{Extra contributions to the covariance}
\label{sec:extra-contr-covar}

Below we will estimate the covariance of the measured shear $\gamma$
when the source galaxies are located at different (unknown) redshifts.
For the purpose, we assume that the redshift distribution of galaxies
is known. In particular, we write the probability to find one galaxy
in an area $\diff^2 \theta$ of the sky with redshift between $z$ and
$z + \diff z$ as
\begin{equation}
  \label{eq:5}
  P_1 = \rho p(z) \, \diff^2 \theta \, \diff z \; .
\end{equation}
Here we model the galaxy angular distribution in terms of a constant
galaxy density $\rho$ on the lens plane. The probability to find two
galaxies, one in a patch $\diff^2 \theta^{(1)}$ of the sky with
redshift in the range $\bigl[ z^{(1)}, z^{(1)} + \diff z^{(1)}
\bigr]$, and one in a  patch $\diff^2 \theta^{(2)}$ with redshift in
the range $\bigl[ z^{(2)}, z^{(2)} + \diff z^{(2)} \bigr]$, can be
written as 
\begin{equation}
  \label{eq:6}
  P_2 = \bigl[ \rho^2 p\bigl( z^{(1)} \bigr) p\bigl( z^{(2)} \bigr) \,
  \diff^2 \theta^{(1)} \, \diff z^{(1)} \, \diff^2 \theta^{(2)} \, 
  \diff z^{(2)} \bigr] \bigr[ 1 + \xi(r) \bigr] \; .
\end{equation}
Here $\xi(r)$ is the two-point correlation function of galaxies at
mutual (3D) distance $r$.  For distances smaller than $10 \, h^{-1}
\mbox{ Mpc}$, this function can be well approximated as a power law
(see Peebles 1993)
\begin{equation}
  \label{eq:7}
  \xi(r) = \left( \frac{5.4 \, h^{-1} \mbox{ Mpc}}{r} \right)^{1.77}
  \; ,
\end{equation}
where $h = H_0 / (100 \mbox{ km s}^{-1} \mbox{ Mpc}^{-1})$ is the
reduced Hubble constant.

We will now evaluate $\Cov(\gamma)$ in the weak lensing limit. In this
approximation, the observed ellipticities are related to the unlensed
ones by the relation
\begin{equation}
  \label{eq:8}
  \epsilon_i = \epsilon^\mathrm{s}_i - \gamma_{0i}(\vec \theta) w(z)
  \; , 
\end{equation}
and thus the expected covariance on $\epsilon$ is given by
\begin{equation}
  \label{eq:9}
  \Cov_{ij}(\epsilon) = \Cov_{ij}(\epsilon^\mathrm{s}) + \Var(w)
  \gamma_{0i}(\vec \theta) \gamma_{0j}(\vec \theta) = c
  \delta_{ij} + \Var(w) \gamma_{0i}(\vec \theta) \gamma_{0j}(\vec
  \theta) \; . 
\end{equation}
A plot of $\Var(w)$ as a function of the lens redshift for a ``typical''
redshift distribution $p(z)$ (see Eq.~\eqref{eq:14} below) is shown in
Fig.~\ref{Fig:var_wz}.

A simple estimator of the shear is given by (see Seitz \& Schneider
1997)
\begin{equation}
  \label{eq:10}
  \gamma(\vec \theta) = - \frac{1}{\langle w \rangle_z} \sum_{n=1}^N
  \epsilon^{(n)} W \bigl(\vec\theta, \vec\theta^{(n)} \bigr) \; .
\end{equation}
Since the weight function $W$ is assumed to be normalized following
Eq.~\eqref{eq:11}, this is an \textit{unbiased\/} estimator. In fact
(for the procedure of ``spatial averaging'' see Paper~II)
\begin{equation}
  \label{eq:12}
  \langle \gamma \rangle (\vec\theta) = - \frac{1}{\langle w
  \rangle_z} \sum_{n=1}^N \langle \epsilon^{(n)}_i \rangle
  W \bigl(\vec\theta, \vec\theta^{(n)} \bigr) \simeq \rho \int_\Omega
  W(\vec\theta, \vec\theta') \gamma_0(\vec\theta') \, \diff^2 \theta'
  \; .
\end{equation}
Thus we recover one simple result already discussed in Paper~II: the
measured shear is the smoothing of the true shear with the spatial
weight function $W$.

Turning to the covariance of $\gamma$, we find
\begin{equation}
  \label{eq:13}
  \begin{split}
    \Cov_{ij}(\gamma; \vec\theta, \vec\theta') = \frac{1}{\langle
    w \rangle_z^2} \biggl\langle \sum_{n,m} & \bigl[ \langle w \rangle_z
    \gamma_{0i}\bigl( \vec\theta^{(n)} \bigr) - \epsilon^{(n)}_i
    \bigr] W \bigl( \vec\theta, \vec\theta^{(n)} \bigr) \times {} \\
    &\bigl[ \langle w \rangle_z \gamma_{0j}\bigl( \vec\theta^{(m)}
      \bigr) - \epsilon^{(m)}_j \bigr] W \bigl( \vec\theta',
      \vec\theta^{(m)} \bigr) \biggr\rangle \; .
  \end{split}
\end{equation}
Inserting here Eq.~\eqref{eq:8}, we find a long expression composed of
three types of terms: terms of the form $\bigl\langle
\epsilon^{\mathrm{s}(n)}_i \epsilon^{\mathrm{s}(m)}_j \bigr\rangle$,
terms of the form $\bigl\langle \epsilon^{\mathrm{s}(n)}_i
\bigr\rangle$, and terms independent of $\epsilon^{\mathrm{s}}$.
Because of the isotropy hypothesis, the first contribution vanishes
unless $n = m$ and $i = j$, and the second always vanishes. Thus we
finally find
\begin{equation}
  \label{eq:15}
  \begin{split}
    \Cov_{ij}(\gamma; \vec\theta, \vec\theta') = {} & \frac{1}{\langle
      w \rangle_z^2} c \delta_{ij} \sum_n W \bigl( \vec\theta,
    \vec\theta^{(n)} \bigr) W \bigl( \vec\theta', \vec\theta^{(n)}
    \bigr) + \\
    & \frac{1}{\langle w \rangle_z^2} \biggl\langle \sum_{n,m} \bigl[
    w \bigl( z^{(n)} \bigr) - \langle w \rangle_z \bigr] \gamma_{0i}
    \bigl( \vec\theta^{(n)} \bigr) W \bigl( \vec\theta,
    \vec\theta^{(n)} \bigr) \times {} \\
    & \phantom{\displaystyle{\frac{1}{\langle w \rangle_z^2}
      \biggl\langle \sum_{n,m}}} \bigl[ w \bigl( z^{(m)} \bigr) - \langle w
    \rangle_z \bigr] \gamma_{0j} \bigl( \vec\theta^{(m)} \bigr) W
    \bigl( \vec\theta', \vec\theta^{(m)} \bigr) \biggr\rangle \; .
  \end{split}
\end{equation}
This equation is composed of two terms. The first is related to the
intrinsic spread of source ellipticities, and in fact it is
proportional to $c = \bigl\langle | \epsilon^\mathrm{s} |^2
\bigr\rangle / 2$. The second term (second and third lines in
Eq.~\eqref{eq:15}) is instead related to the spread in redshift of
galaxies. In order to obtain a more general expression, we adopt a
technique already used in Paper~II and average the expression for
$\Cov(\gamma)$ over the source positions $\bigl\{ \vec\theta^{(n)}
\bigr\}$ and redshifts $\bigl\{ z^{(n)} \bigr\}$. As a result, the
first term becomes simply
\begin{equation}
  \label{eq:16}
  \Gamma_1 = 
  \frac{1}{\langle w \rangle_z^2} c \delta_{ij} \sum_n W \bigl(
  \vec\theta, \vec\theta^{(n)} \bigr) W \bigl( \vec\theta',
  \vec\theta^{(n)} \bigr) \longmapsto \frac{1}{\langle w \rangle_z^2}
  c \delta_{ij} \rho \int_\Omega W(\vec\theta, \vec\theta'')
  W(\vec\theta', \vec\theta'') \, \diff^2 \theta'' \; .
\end{equation}
This term has already been studied in detail in Paper~II.

The second term of Eq.~\eqref{eq:15} deserves a more detailed
discussion. In order to carry out the average over the source
redshifts, we need to split the sum over $n$ and $m$ in two sums, one
involving the terms for which $n = m$, and the other involving the
terms $n \neq m$. For the first terms we have
\begin{multline}
  \label{eq:17}
  \Gamma_2 = 
  \frac{1}{\langle w \rangle_z^2} \biggl\langle \sum_n \bigl[ w \bigl(
  z^{(n)} \bigr) - \langle w \rangle_z \bigr]^2 \gamma_{0i} \bigl(
  \vec\theta^{(n)} \bigr) \gamma_{0j} \bigl( \vec\theta^{(n)} \bigr) W
  \bigl( \vec\theta, \vec\theta^{(n)} \bigr) W \bigl(
  \vec\theta', \vec\theta^{(n)} \bigr) \biggr\rangle \longmapsto \\
  \longmapsto \frac{1}{\langle w \rangle_z^2} \rho \Var(w) \int_\Omega
  \gamma_{0i} (\vec\theta'') \gamma_{0j} (\vec\theta'') W(\vec\theta,
  \vec\theta'') W(\vec\theta', \vec\theta'') \, \diff^2 \theta'' \; .
\end{multline}
Note that this term is related to single galaxies, and thus does not
involve the correlation function $\xi(r)$. On the other hand, for
pairs of galaxies (i.e.\ if $n \neq m$ in the last term of
Eq.~\eqref{eq:15}) we find a rather complicated expression
\begin{multline}
  \label{eq:18}
  \Gamma_3 = 
  \frac{1}{\langle w \rangle_z^2} \biggl\langle \sum_{n \neq m} \bigl[
  w \bigl( z^{(n)} \bigr) - \langle w \rangle_z \bigr] \gamma_{0i}
  \bigl( \vec\theta^{(n)} \bigr) W \bigl( \vec\theta,
  \vec\theta^{(n)} \bigr) \bigl[ w \bigl( z^{(m)} \bigr) - \langle w
  \rangle_z \bigr] \gamma_{0j} \bigl( \vec\theta^{(m)} \bigr) W \bigl(
  \vec\theta', \vec\theta^{(m)} \bigr) \biggr\rangle \longmapsto \\
  \longmapsto \frac{1}{\langle w \rangle_z^2} \rho^2 \int_0^\infty
  \!\! p(z'') \, \diff z'' \int_\Omega \! \diff^2 \theta'' \int_0^\infty
  \!\! p(z''') \, \diff z''' \int_\Omega \! \diff^2 \theta''' 
  \bigl[ w(z'') - \langle w \rangle_z \bigr] \gamma_{0i}(\vec\theta'')
  W(\vec\theta, \vec\theta'') \times {} \\
  \bigl[ w(z''') - \langle w \rangle_z \bigr]
  \gamma_{0j}(\vec\theta''')  W(\vec\theta', \vec\theta''') \xi(r) \; .
\end{multline}
In this expression $r$ is the distance between two galaxies at
redshifts $z''$ and $z'''$ and angular positions $\vec\theta''$ and
$\vec\theta'''$.

The covariance of $\gamma$ is the sum of the terms in
Eqs.~(\ref{eq:16}--\ref{eq:18}), each of which, as discussed above,
arises from a different source of noise. In summary,
\begin{itemize}
\item The contribution of Eq.~\eqref{eq:16} is related to the spread
  in the intrinsic ellipticities of source galaxies (parameterized by
  $c$). It is proportional to $1/\rho$ (see the normalization
  condition \eqref{eq:11}) and has a correlation length
  proportional to the correlation length of $W$. Note that this term
  is proportional to $\delta_{ij}$, so that no correlation between
  $\gamma_1$ and $\gamma_2$ arises.
\item The contribution of Eq.~\eqref{eq:17} is related to the spread
  in redshift of the sources (parameterized by $\Var(w)$).  It is
  proportional to $1/\rho$, has the same correlation length of the
  previous term, and, in addition, is proportional to $\gamma_0^2$. This
  term is not diagonal.
\item The contribution of Eq.~\eqref{eq:18} is due to the correlation
  between galaxies (described by the two-point correlation function
  $\xi(r)$).  It is independent of the density of galaxies (provided
  $\xi(r)$ does not depend on $\rho$; see comments by Peebles 1993
  after Eq.~(19.39)), has a correlation length somewhat longer than
  that of the other terms and is proportional to $\gamma_0^2$. This
  term is not diagonal.
\end{itemize}
A complete description of the effect of the first term has already
been provided in Paper~II. Its effect for the detection of a double
lens was analyzed in Sect.~\ref{sec:expect-vari-nabla}. 

The order of magnitude of the ratio between the term of
Eq.~\eqref{eq:17} and the one of Eq.~\eqref{eq:16} can be argued to be
\begin{equation}
  \label{eq:19}
  \frac{\Gamma_2}{\Gamma_1} \simeq \frac{\Var(w) \gamma_0^2}{c} \; .
\end{equation}
Typical values for $c$ are $c \simeq 0.02$, and typical values for
$\Var(w)$ can be estimated from Fig.~\ref{Fig:var_wz}. Thus, if we
refer to a single lens at redshift $z_\mathrm{d} = 0.2$ with a shear
of the order of $\gamma \simeq 0.3$, we find $\Gamma_2 / \Gamma_1
\simeq 0.2$.  Note that, since calculations here are performed in the
weak lensing approximation, the shear cannot be much larger than
$0.3$, and thus the ratio obtained should be taken as an
(approximate) upper limit. In any case, the effects of this term will
be included in the simulations that will be described in Sect.~6.2.

As to the third term $\Gamma_3$, it is clearly very complex, both in
relation to its mathematical structure and to the physical ingredients
involved. Therefore, instead of attempting here an order of magnitude
estimate for it, we prefer to postpone a complete discussion of its
impact to a future paper that will make use of suitable simulations in
the cosmological context.

\subsubsection{Bias on the reduced shear}
\label{sec:bias-reduced-shear}

Another effect encountered when source galaxies are ``placed'' at
different redshifts is a bias on $\nabla \wedge \vec u$, which is not
bound to vanish any more. This effect could, in principle, be
misinterpreted as the signature of a double lens.  Do we have to worry
about this contribution?  Note that if this effect turned out to be
significant, it would have an important impact on all single lens mass
reconstruction methods, because these are all based on the curl-free
character of the shear induced by gravitational lensing.

In order to be more specific about the origin of this effect, we start
from some well-known results (Seitz \& Schneider 1997; see also
Paper~III). Let us consider a sample of galaxies with redshift
distribution $p(z)$, behind a single cluster at redshift
$z_\mathrm{d}$. As the critical density $\Sigma_\mathrm{c}(z)$ depends
on the source redshift, the lens acts with differential strength on
the lensed galaxies (see Sect.~2.1).  The cosmological weight function
$w(z)$ controls the ray-tracing transformation. In particular, the
(redshift dependent) reduced shear takes on the form $g(\vec\theta, z)
= w(z) \gamma(\vec\theta) / \bigl[ 1 - w(z) \kappa(\vec\theta)
\bigr]$. The fact that $w(z)$ enters the expression of $g(\vec\theta,
z)$ nonlinearly complicates the lensing analysis. Note that even
though the relevant ray-tracing transformation for a single lens is
symmetric (e.g., see Eq.~(9) of Paper~III), because of the spread in
redshift distribution the arguments generally used to identify a
curl-free field $\vec{\tilde u}$ do not hold any more. In fact, if we
observe a single lens and just ignore the redshift distribution of the
sources, we would naively start by identifying the observed average
ellipticity with the reduced shear, using the relation $\langle
\epsilon \rangle = -g$. This mistake would lead in general to a
quantity $\vec{\tilde u}$, defined by Eq.~\eqref{eq:u_tilde}, with non
vanishing curl.

For subcritical lenses Seitz \& Schneider (1997) have shown that an
accurate, but approximate, lensing analysis can be carried out
\textit{as if\/} all galaxies were at a single redshift, provided the
reduced shear $g(\vec\theta)$ in Eq.~\eqref{eq:u_tilde} is evaluated
from the average ellipticity following the relation $\langle \epsilon
\rangle \langle w^2 \rangle_z / \langle w \rangle_z^2 = -g$.
Correspondingly, the dimensionless density distribution obtained from
the lens reconstruction changes by a factor $\langle w^2 \rangle_z /
\langle w \rangle_z$. If the above corrected relation between average
ellipticity and reduced shear entering the definiton of $\vec{\tilde
  u}$ is kept into account, then, in this approximate description, the
``spurious curl'' (that would be associated with the naive use of
$\langle \epsilon \rangle = -g$) is eliminated.

In order to evaluate the magnitude of the effects (described in this
subsection) associated with the spread in the distances of the source
galaxies, we have carried out simulations that will be described in
the next section. Here we summarize the main results. Using a
reasonable redshift distribution for the sources, and for a wide range
of redshifts $z_\mathrm{d}$ of the single deflector, we find an
extremely small value of $\nabla \wedge \vec{\tilde u}$. For example,
for a nearly critical single lens at redshift $z_\mathrm{d} = 0.3$ the
value of the relevant curl is about $50$ times smaller than the value
associated with the double lens effect addressed in this paper. The
nearly critical condition adopted in this example certainly
overemphasizes the signal, with respect to the case of weaker lenses,
and is unfavorable to the approximation suggested by Seitz \&
Schneider. Still, if we apply the simple correction factor $\langle
w^2 \rangle_z / \langle w \rangle_z^2 $ described above, we gain an
extra factor of $4$. We can thus state that the effect of the redshift
distribution discussed in this subsection, in the example considered,
is contained to be about $200$ times smaller than the effect of double
lensing we are interested in. From this we conclude that the simple
arguments and derivations provided in the main part of the paper are
not affected by the spread in the distances of the source galaxies
significantly.


\section{Simulations}
\label{sect:simulations}

The main goal of this section is to check the analytical framework
developed in the paper and to provide convincing evidence that effects
characteristic of double lensing are within reach of the observations.
For this reason we follow two types of investigation: a
semi-analytical approach (Sect.~6.1), where we test the applicability
of our asymptotic analysis, and a study (Sect.~6.2) that may be seen
as a numerical observation, where we simulate the distorsions of a set
of source galaxies in a given field under realistic conditions (see
below) and we perform the statistical lensing analysis on the observed
ellipticities. For simplicity, the simulations are carried out under
the assumption that no clustering is present in the source population
(Eq.~\eqref{eq:6} with $\xi(r) = 0$).

\subsection{Check of the analytical framework}

A first test is based on the following steps of a semi-analytical
approach:

\begin{itemize}

\item The population of source galaxies is taken to be at a single
  redshift ($z_\mathrm{s} = 2.0$).

\item A simple model of a double lens is chosen. For simplicity, two
  centrally-symmetric projected mass distributions are used, within a
  favorable geometric configuration (see Sect.~4.1).
  
\item The two clusters are taken to be located at redshift $z^{(1)} =
  0.1$ and $z^{(2)} = 0.4$. Since the source population and the double
  lens we have in mind have such large distances, the values of the
  cosmological parameters $\Omega$ and $\Omega_\Lambda$ need to be
  specified. In the following we refer to the case $\Omega = 0.3$,
  $\Omega_\Lambda = 0.7$; moreover, we adopt $H_0 = 65
  \,$km$\,$s$^{-1}\,$Mpc$^{-1}$.
  
\item The two functions $\vec\beta^{(1)}(\vec\theta)$ and
  $\vec\beta^{(2)}(\vec\theta)$ are calculated. Thus the ray-tracing
  function $\vec\theta^\mathrm s(\vec\theta)$ is derived (see
  Eq.~\eqref{eq:ray-tracing}).
  
\item The Jacobian matrix $A_0(\vec\theta)$ is calculated from the map
  $\vec\theta^\mathrm s(\vec\theta)$. This allows us to calculate the
  various quantities defined in Eqs.~(\ref{eq:B18}--\ref{eq:B22}). In
  the following, these are taken to hold as definitions also for
  strong lenses.
  
\item The ``effective'' Jacobian matrix $A_\mathrm{s}(\vec\theta)$ is
  obtained from $A_0(\vec\theta)$ using Eq.~\eqref{eq:+-f}.

\item From $A_\mathrm{s}(\vec\theta)$, the reduced shear
  $g(\vec\theta)$ and the field $\vec{\tilde u}(\vec\theta)$ are
  calculated.
  
\item The divergence and the curl of $\vec{\tilde u}$ are compared
  with the Laplacians of $\ln(\sigma)$ and $\tau$ (Eqs.~\eqref{eq:B26}
  and \eqref{eq:B27}).

\end{itemize}

Note that, by following this procedure, we are considering the
``effective'' Jacobian matrix $A_\mathrm{s}$ {\it without errors}. We
do so, because errors on $A_\mathrm{s}$ are discussed separately (see
next Subsection 6.2, and Paper~I and II) and because we are focusing
here mostly on a test of the adequacy of Eqs.~\eqref{eq:B26} and
\eqref{eq:B27}. Let us now describe some of the steps in further
detail.

The two clusters, taken to be located at redshift $z^{(1)} = 0.1$ and
$z^{(2)} = 0.4$ (so that the value of the parameter $\Delta$ defined
in Eq.~\eqref{eq:Delta} is $\Delta \simeq 0.794$ in the assumed
cosmological model), have been modeled using the following profiles
(close to isothermal profiles) for the projected density (see
Schneider, Ehlers, \& Falco 1992):
\begin{equation}
  \kappa(\vec\theta) = {\bigl[ 1 + (\theta / \theta_\mathrm{c})^2 / 2
  \bigr] \kappa_\mathrm{c} \over \bigl[ 1 + (\theta /
  \theta_\mathrm{c})^2 \bigr]^{3/2}} \; ,
\end{equation}
with $\theta = \| \vec\theta \|$. Here $\kappa_\mathrm{c} = \kappa(
\vec{0} )$ is the maximum value of the projected density, and
$\theta_\mathrm{c}$ is a length scale (core radius). The deflection
$\vec\beta(\vec\theta)$ for such profiles takes a particularly simple
form:
\begin{equation}
  \vec\beta(\vec\theta) = {\kappa_\mathrm{c} \over \sqrt{ 1 + (\theta
  / \theta_\mathrm{c})^2 }} \vec\theta\; .
\end{equation}
The intrinsic \textit{physical\/} parameters of the two clusters, i.e.
the core radius in Mpc and central density in kg$\,$m$^{-2}$
($1\,$kg$\,$m$^{-2} \simeq 478 M_\odot\,$pc$^{-2}$) have been assumed
to be equal. For simplicity, the first cluster is centered at
$\vec\theta = (5', 3'\,30'')$ and the second cluster at $\vec\theta =
(5', 6'\,30'')$. In addition, we have considered two cases: a case of
``weak'' lenses, for which the second order expansion considered in
Sect.~3 is expected to provide accurate results, and a case of
``strong'' lenses, with ``joint mass density'' $1 - \sigma$ too high
in relation to the asymptotic expansion. For the strong case the mass
associated within our field of $10' \times 10'$ for the far cluster is
$M \simeq 3.14 \times 10^{15} \, M_\odot$.  The latter case will show
that, at least qualitatively, the analysis developed in this paper is
applicable even to relatively strong lenses. The cluster parameters
adopted for the tests presented here are summarized in Table~1.

In Figs.~\ref{Fig:light2} and \ref{Fig:density2} we display the
combined luminosity and dimensionless density distributions.

\begin{table}[!b]
  \caption{The parameters used for the simulations.}
  \small
  \begin{tabular}{l | c c c c c c}
    \hline
    \hline
    \multicolumn{2}{c}{component} & redshift &
    \multicolumn{2}{c}{density} & \multicolumn{2}{c}{core radius}\\
    \multicolumn{2}{c}{} & $z$ & $\kappa_\mathrm{c}$ &
    $\kappa_\mathrm{c} \Sigma_\mathrm{c}$ (kg$\,$m$^{-2}$) &
    $\theta_\mathrm{c}$ (angle) & $r_\mathrm{c}$ (Mpc) \\
    \hline
    & cl 1 & $0.1$ & $0.054$ & 0.5 & $2'\,31''$ & $0.3$ \\
    \raisebox{2.0ex}[0pt]{weak} & \#2 & $0.4$ & $0.120$ & 0.5 &
    $0'\,52''$ & $0.3$ \\ 
    \hline
    & cl 2 & $0.1$ & $0.401$ & 3.7 & $2\,31''$ & $0.3$ \\
    \raisebox{2.0ex}[0pt]{strong} & \#2 & $0.4$ & $0.891$ & 3.7 &
    $0'\,52''$ & $0.3$ \\
    \hline
    \hline
  \end{tabular}
\end{table}

The second step has been the calculation of the ray-tracing function
$\vec\theta^\mathrm s(\vec\theta)$. This function has been calculated
on a grid of $100 \times 100$ points, representing a square with a
side of $10'$. The Jacobian matrix has been calculated on the same
grid using the approximate derivatives determined from the differences
of $\vec\theta^\mathrm s$ in neighboring points.

The plot of the semi-trace $\sigma$ of $A_0$ (see Eq.~\eqref{eq:B19})
presents some interesting features. We first note that for weak lenses
$\sigma = 1 - \kappa^{(1)} - \kappa^{(2)}$. However, when the lens is
strong, a curious effect is noted (cfr.\
Fig.~\ref{Fig:trA0-density2}): the maximum corresponding to the mass
distribution of the second cluster is significantly smoothed and
lowered. This has to do with the coupling of the two deflectors.  The
effect can be broadly described as a {\it magnification\/} of the mass
distribution of the second deflector because of the lensing effect of
the first cluster. In fact, while the value of $\kappa^{(1)} +
\kappa^{(2)}$ exceeds unity (see Fig.~\ref{Fig:density2}), the double
lens remains sub-critical with $1 - \sigma$ below unity.

Proceeding along the list of steps indicated above, we have calculated
the reduced shear $g(\vec\theta)$ and the field $\vec{\tilde
  u}(\vec\theta)$. At the end of the process, the two quantities
$\nabla \cdot \vec{\tilde u}$ and $\nabla \wedge \vec{\tilde u}$ are
available on a $100 \times 100$ grid. Note that all differentiations
have been performed using a 3-point, Lagrangian interpolation.

The results of our tests for the weak case are summarized in
Figs.~\ref{Fig:divu-lapsigma1}--\ref{Fig:curlu-laptau_1}. Within the
numerical errors, we find that Eqs.~\eqref{eq:B26} and \eqref{eq:B27}
are verified. Of course, since in this case the magnitude of the
effect is very small (the maximum value of $\tau$ is approximately $2
\times 10^{-4}$), we should not expect that double lensing of this
type could be actually detected. In passing we note that the order of
magnitude for the various quantities shown in the figures is
consistent with this value of $\tau$.

The tests for the strong case (see
Figs.~\ref{Fig:divu-lapsigma2}--\ref{Fig:curlu-laptau_2}) are rather
surprising. In fact, Fig.~\ref{Fig:divu-lapsigma2} shows that
Eq.~\eqref{eq:B26} holds approximately also in this case. However,
Eq.~\eqref{eq:B27} turns out to be inadequate, as shown by
Fig.~\ref{Fig:curlu-laptau2}. This behavior reflects the fact that the
lens configuration here has a rather high value of $1 - \sigma$ (its
maximum is about $0.928$), and a small value of $\tau$ (maximum about
$0.01$). A heuristic way to correct Eq.~\eqref{eq:B27} is the
following:
\begin{equation}
  \label{eq:heuristic}
  \nabla \wedge \vec{\tilde u} = \nabla \cdot \left( {\nabla \tau
  \over \sigma} \right) \; .
\end{equation}
The use of $\sigma$ in the denominator of the previous equation is
suggested by analogy with Eq.~\eqref{eq:B26}, where $\nabla (\ln
\sigma) = (\nabla \sigma) / \sigma$ is used. Moreover, we note that,
to second order, Eq.~\eqref{eq:heuristic} is identical to
Eq.~\eqref{eq:B27}. The new equation turns out to be a better
approximation than Eq.~\eqref{eq:B27}, especially for regimes away
from the strong and the weak limiting cases considered above; in such
intermediate cases Eq.~\eqref{eq:heuristic} performs better, typically
by a factor of $2$.

Finally, we have considered the problem of the dark cluster (see
Sect.~4.4). In this case the simplified simulations in the
semi-analytical approach described above have been performed with the
aim of demonstrating how the redshift of a dark cluster can be
determined when the redshift and the mass distribution of the luminous
cluster are known. In order to find the unknown redshift we have
minimized $S$, the integral of the square of the left hand side of
Eq.~\eqref{eq:findz1} or of Eq.~\eqref{eq:findz2}, depending on
whether the dark cluster is considered to be closer to us than the
luminous cluster or not. Simulations have been performed using the
parameters of the strong lens configuration, by taking either the
first or the second cluster as the dark cluster.
Figure~\ref{Fig:findz-0.1} shows $S$ for the case where the dark
cluster is in the front, while Fig.~\ref{Fig:findz-0.4} shows the same
quantity for the case where the dark cluster is behind.  Note that $S$
has a minimum (near or equal to $0$) for the true redshift of the dark
cluster. Note also that the second minimum of $S$ occurs where the
condition $\Delta = 0$ is met: in this case the quantities
$\vec\zeta^{(1)}$, $\vec\zeta^{(2)}$, and $\vec\zeta^\mathrm{s}$ all
reduce to $\vec\theta$. We should stress that if a dark cluster is
detected from a double lens signature, as described in this paper, and
if the key characteristics of the bright cluster are known from
independent diagnostics, then \textit{the reconstruction method
  developed here leads to the full determination of both the distance
  and the mass distribution of the invisible cluster}.

\subsection{More realistic simulations}

In order to ascertain the relevance of our analysis with respect to
present and future observations, we have performed a set of
simulations described in the following steps:

\begin{itemize}
\item We generate $N$ source galaxies.  Each galaxy is described by
  its location in the sky, its redshift, and its source ellipticity.
  Locations are chosen following a uniform distribution on the
  observed locations $\vec\theta$ (thus neglecting the magnification
  effect).  The redshift of source galaxies has been drawn from a
  gamma distribution used also by other authors (e.g., Brainerd et
  al.\ 1996)
  \begin{equation}
    \label{eq:14}
    p(z) = \frac{z^2}{2 z_0^3} \exp\bigl[ -(z/z_0) \bigr] \; ,
  \end{equation}
  with $z_0 = 1/3$ (other values of $z_0$ have also been tested).
  Finally, ellipticities $\epsilon^\mathrm{s}$ are taken from a
  truncated Gaussian distribution (see Eq.~(32) of Paper III)
  characterized by $c = 0.02$.  This value is probably unrealistically
  large; it is adopted in order to test our framework under relatively
  unfavorable conditions. 
\item As a model for the double lens we use the ``strong'' double lens
  model introduced in the previous subsection, within the same field
  of view of $10' \times 10'$.
\item Source ellipticities $\epsilon^\mathrm{s}$ are transformed into
  observed ellipticities $\epsilon$ using the true asymmetric Jacobian
  matrix associated with the lens. Note that the redshift of each
  galaxy has been taken into account when performing this
  transformation.
\item The statistical lensing reconstruction procedure is applied to
  the observed ellipticities. The reduced shear map $g(\vec\theta)$ is
  derived by averaging observed ellipticities of nearby galaxies, from
  the relation $\langle \epsilon \rangle = -g$ properly corrected for
  the redshift distribution.
\item From the shear map we calculate the other relevant quantities:
  the vector field $\vec {\tilde u}(\vec\theta)$, its divergence, and
  its curl.
\item We then look for significant peaks in the curl map, which would
  reveal the double nature of the lens.
\end{itemize}

Several simulations have been performed with different source galaxy
densities. The results obtained are very encouraging, since a
significant signal in the the curl of $\vec {\tilde u}$ is already
detected with a galaxy density of $100 \mbox{ gal arcmin}^{-2}$. Such
relatively high density has already been reached by current
observations.  For example, new observations with the Very Large
Telescope of the cluster MS1008${}.{}$1$-$1224 come close to it
(Lombardi et al.\ 2000, Athreya et al.\ 2000; see also Hoekstra et
al.\ 2000 for observations of MS1054 with the HST).

Figure~\ref{Fig:dens_plot} shows a typical result obtained in our
simulations.  The figure shows the density plot of the observed
$\nabla \wedge \vec{\tilde u}$ and, superimposed, the contour plot of
$\nabla \wedge \vec{\tilde u}_0$, i.e.\ the true quantity that we
should measure in the limit of infinite galaxy density. For both plots
we have used the same smoothing spatial weight function, a Gaussian of
width $\sigma_\mathrm{W} = 30''$ (see Sect.~5.2). We thus conclude
that, given the conditions adopted in our simulations, the
characteristic signature of double lensing should be within reach of
current observations.

Finally, in order to test the reliability of the detection, we have
performed analogous simulations using a single cluster. In this case,
as explained above, we expect a (nearly) vanishing $\nabla \wedge
\vec{\tilde u}$, especially after the correction for the spread of
background sources is applied (see Sect.~5.3.2).
Figure~\ref{Fig:density-single} shows a map of this quantity (with the
correction applied) in the case of a single lens at redshift $0.1$
(the parameters of this lens are the same as those of the first lens
considered in the double configuration). Note that the noise observed
is well below the signal shown in Figure~\ref{Fig:dens_plot}.


\section{Conclusions}

In this paper we have generalized the statistical analysis at the
basis of mass reconstructions of weak gravitational lenses from the
standard case of single lenses to the case of double lenses. From one
point of view, this study leads to practical tools to deal with
``errors'' that might occur when the standard single lens analyses are
applied to the not infrequent cases where two massive clusters happen
to be partially aligned along the line of sight. In reality, our study
of double lenses, based on the analytical framework mentioned in the
first item below then checked and extended by means of simulations,
has opened the way to a number of interesting results. This is a list
of the main points made in our paper:

\begin{enumerate}
\item A consistent analytical framework has been constructed where the
  contribution of the small asymmetry in the Jacobian matrix induced
  by double lensing is retained to two orders in the weak lensing
  asymptotic expansion.
\item Given the known properties of the distribution of (bright)
  clusters of galaxies, we have shown that a few configurations are
  likely to be present in the sky, for which the small effects
  characteristic of double lensing may already be within detection
  limits.
\item As a separate astrophysical application, we have demonstrated
  that, if the characteristic signature of double lensing appears in
  an observed configuration where a single bright cluster exists, with
  properties well constrained by independent diagnostics, the location
  and the mass distribution of the dark cluster, if present and
  responsible for the effect, can in principle be reconstructed
  unambiguously.
\item We have checked that the redshift distribution of the source
  galaxies should not confuse the signature of double lensing.
\item An examination of the relevant contributions to the noise of the
  shear measurements has brought up a limitation of weak lensing
  analyses related to the clustering of source galaxies. In
  particular, weak lensing studies of single clusters are found to be
  characterized by a lower limit for the expected noise, regardless of
  the depth of the images and of the density of source galaxies. A
  firm quantitative estimate of this surprising effect will be
  provided an a separate article.
\end{enumerate}

\acknowledgements This work was partially supported by MURST of Italy.
We thank Peter Schneider for several stimulating discussions that have
helped us improve this paper significantly.


\begin{appendix}

\section{Second order weak lensing analysis}

In this Appendix we derive Eqs.~\eqref{eq:B26} and \eqref{eq:B27}.

We start from the definition of $\vec{\tilde u}$, which, to second
order, is
\begin{equation}
  \begin{split}
    \vec{\tilde u} &= 
    \begin{pmatrix}
      2 - \sigma + \gamma_1 & \gamma_2 \\
      \gamma_2 & 2 - \sigma - \gamma_1
    \end{pmatrix}
    \begin{pmatrix}
      \gamma_{1,1} + \gamma_{2,2} \\ 
      \gamma_{2,1} - \gamma_{1,2}
    \end{pmatrix} - \\
    & 
    \begin{pmatrix}
      \sigma_{,1} &  \sigma_{,2} \\ 
      \sigma_{,2} & -\sigma_{,1}
    \end{pmatrix}
    \begin{pmatrix}
      \gamma_1 \\
      \gamma_2
    \end{pmatrix} \; .
  \end{split}
\end{equation}
We recall that $\sigma = 1 + \mathcal{O}(\epsilon)$. By inserting
Eqs.~(\ref{eq:B19}--\ref{eq:B22}) here, rather long expressions are
found:
\begin{align}
  \label{eq:u-2_1} 
  \tilde u_1 =& \frac{1}{4} \Bigl[ 4 \nabla^2 \theta^\mathrm{s}_1 -
  \theta^\mathrm{s}_{1,11} \nabla \cdot \vec\theta^\mathrm{s} - 2
  \theta^\mathrm{s}_{1,22} \theta^\mathrm{s}_{2,2} + \nonumber\\ 
  & \theta^\mathrm{s}_{2,12} \bigl( \theta^\mathrm{s}_{2,2} -
  \theta^\mathrm{s}_{1,1} \bigr) + \bigl( \theta^\mathrm{s}_{1,2} +
  \theta^\mathrm{s}_{2,1} \bigr) \nabla \wedge
  \vec\theta^\mathrm{s}_{,1} \Bigr] \; , \\
  \label{eq:u-2_2}
  \tilde u_2 =& \frac{1}{4} \Bigl[ 4 \nabla^2 \theta^\mathrm{s}_2 -
  \theta^\mathrm{s}_{2,22} \nabla \cdot \vec\theta^\mathrm{s} - 2
  \theta^\mathrm{s}_{2,11} \theta^\mathrm{s}_{1,1} - \nonumber\\ 
  & \theta^\mathrm{s}_{1,12} \bigl( \theta^\mathrm{s}_{2,2} -
  \theta^\mathrm{s}_{1,1} \bigr) - \bigl(\theta^\mathrm{s}_{1,2} +
  \theta^\mathrm{s}_{2,1} \bigr) \nabla \wedge
  \vec\theta^\mathrm{s}_{,2} \Bigr] \; . 
\end{align}
From these expressions we can calculate the divergence and the curl of
$\vec{\tilde u}$. The expressions obtained by a straightforward
application of the definitions contain also terms of order
$\epsilon^3$ or higher, which are to be discarded.

Let us start from the curl of $\vec{\tilde u}$. From the previous
equations, after some manipulations, we obtain
\begin{equation}
  \begin{split}
    \nabla \wedge \vec{\tilde u} = & 
    \nabla^2 (\nabla \wedge \vec\theta^\mathrm{s}) - \frac{1}{2}
    \Bigl[ \theta^\mathrm{s}_{1,1} \nabla \wedge
    \vec\theta^\mathrm{s}_{,11} + \theta^\mathrm{s}_{2,2} \nabla
    \wedge \vec\theta^\mathrm{s}_{,22} + \\
    & \bigl( \theta^\mathrm{s}_{1,2} + \theta^\mathrm{s}_{2,1} \bigr) \nabla
    \wedge \vec\theta^\mathrm{s}_{,12} + \\
    & \theta^\mathrm{s}_{1,11} \nabla \wedge
    \vec\theta^\mathrm{s}_{,1} + \theta^\mathrm{s}_{2,22} \nabla
    \wedge \vec\theta^\mathrm{s}_{,2} + \\
    & \theta^\mathrm{s}_{1,12} \theta^\mathrm{s}_{1,22} -
    \theta^\mathrm{s}_{2,11} \theta^\mathrm{s}_{2,12} \Bigr] \; .
  \end{split}
\end{equation}
The terms in this expression are organized by rows. Recalling that
$\nabla \wedge \vec\theta^\mathrm s = 2 \tau \sim \epsilon^2$, we
recognize that the terms in the second and third lines are of order
$\epsilon^3$, and thus should be dropped. A simple analysis also shows
that the two terms in the last line cancel out. Finally, as to the
first line, we note that $\theta^\mathrm{s}_{1,1}$ and
$\theta^\mathrm{s}_{2,2}$ are of the form $1 + \mathcal{O}(\varepsilon)$.
Thus at the end we obtain
\begin{equation}
  \nabla \wedge \vec{\tilde u} \simeq {1 \over 2} \nabla^2 (\nabla \wedge
  \vec\theta^\mathrm{s}) \; ,
\end{equation}
that is Eq.~\eqref{eq:B27}.

Let us now turn to the divergence of $\vec{\tilde u}$. From
Eqs.~\eqref{eq:u-2_1} and \eqref{eq:u-2_2}, a rather long calculation gives
\begin{equation}
  \begin{split}
    \nabla \cdot \vec{\tilde u} =& \nabla^2 ( \nabla \cdot
    \vec\theta^\mathrm{s} ) + \\
    & \frac{1}{4} \Big[ \nabla \cdot \vec
    \theta^\mathrm{s} \bigl( \theta^\mathrm{s}_{1,122} -
    \theta^\mathrm{s}_{1,111} + \theta^\mathrm{s}_{2,112} -
    \theta^\mathrm{s}_{2,222} \bigr) - \\
    & 4 \theta^\mathrm{s}_{1,1} \theta^\mathrm{s}_{2,112} - 
    4 \theta^\mathrm{s}_{2,2} \theta^\mathrm{s}_{1,122} - \\
    & \nabla^2 \theta^\mathrm{s}_1 \bigl( \theta^\mathrm{s}_{1,11} -
    \theta^\mathrm{s}_{1,22} + 2 \theta^\mathrm{s}_{2,12} \bigr) + \\ 
    & \nabla^2 \theta^\mathrm{s}_2 \bigl( \theta^\mathrm{s}_{2,11} -
    \theta^\mathrm{s}_{2,22} - 2 \theta^\mathrm{s}_{1,12} \bigr) + \\
    & \bigl( \theta^\mathrm{s}_{1,2} + \theta^\mathrm{s}_{2,1} \bigr)
    \bigl( \nabla \wedge \vec\theta^\mathrm{s}_{,11} - \nabla \wedge
    \vec\theta^\mathrm{s}_{,22} \bigr) \Big] \; .    
  \end{split}
\end{equation}
The first line of this expression needs no special explanations, since
we recall that $\nabla \cdot \vec\theta^\mathrm{s} = 2 \sigma$. The
sum of the terms in the second and third lines can be shown to be
equal (dropping terms of order $\epsilon^3$ or higher) to $- 4 \sigma
\nabla^2 \sigma$. By replacing $2 \theta^\mathrm{s}_{2,12}$ by $2
\theta^\mathrm{s}_{1,22}$ and $2 \theta^\mathrm{s}_{1,12}$ by $2
\theta^\mathrm{s}_{2,11}$ (which is allowed to second order), the
terms in parentheses in the fourth and fifth lines become respectively
$\nabla^2 \theta^\mathrm{s}_1$ and $\nabla^2 \theta^\mathrm{s}_2$.
Finally, the last line can be discarded, because it contains two terms
of order $\epsilon^3$. In conclusion, we obtain
\begin{equation}
  \nabla \cdot \vec{\tilde u} \simeq 2 \nabla^2 \sigma - \sigma \nabla^2
  \sigma - \frac{1}{4} \bigl\| \nabla^2 \vec\theta^\mathrm s \bigr\|^2 \; .
  \label{eq:App2}
\end{equation}
This expression can be replaced by $\nabla^2 ( \ln \sigma )$. In fact,
to second order we have
\begin{equation}
  \nabla^2( \ln \sigma ) = (2 - \sigma) \nabla^2 \sigma - \|
  \nabla \sigma \|^2 \; .
  \label{eq:App}
\end{equation}
We now apply the vector identity $\nabla( \nabla \cdot \vec A) =
\nabla^2 \vec A - \nabla \wedge \nabla \wedge \vec A$ (valid for any
vector field $\vec A$) to the last term of Eq.~\eqref{eq:App}:
\begin{equation}
  \begin{split}
    \bigl\| \nabla ( \nabla \cdot \vec\theta^\mathrm{s} ) \bigr\|^2 =&
    \bigl\| \nabla^2 \vec\theta^\mathrm{s} \bigr\|^2 - 2 \bigl(
    \nabla^2 \vec\theta^\mathrm{s} \bigr) \cdot \bigl( \nabla \wedge
    \nabla \wedge \vec\theta^\mathrm{s} \bigr) + \\ 
    & \bigl\| \nabla \wedge \nabla \wedge \vec\theta^\mathrm{s}
    \bigr\|^2 = \bigl\| \nabla^2 \vec\theta^\mathrm{s} \bigr\|^2 \; ,    
  \end{split}
\end{equation}
where the last equality holds because $\nabla \wedge
\vec\theta^\mathrm{s} = 2 \tau$ is of order $\epsilon^2$. Thus
finally we can rewrite Eq.~\eqref{eq:App} as
\begin{equation}
  \nabla^2( \ln \sigma ) = (2 - \sigma) \nabla^2 \sigma - \frac{1}{4}
  \| \nabla^2 \vec\theta^\mathrm{s} \|^2 \; .
\end{equation}
Comparing this expression with Eq.~\eqref{eq:App2} we obtain the
desired relation \eqref{eq:B26}.

\end{appendix}



\begin{center}
  \includegraphics{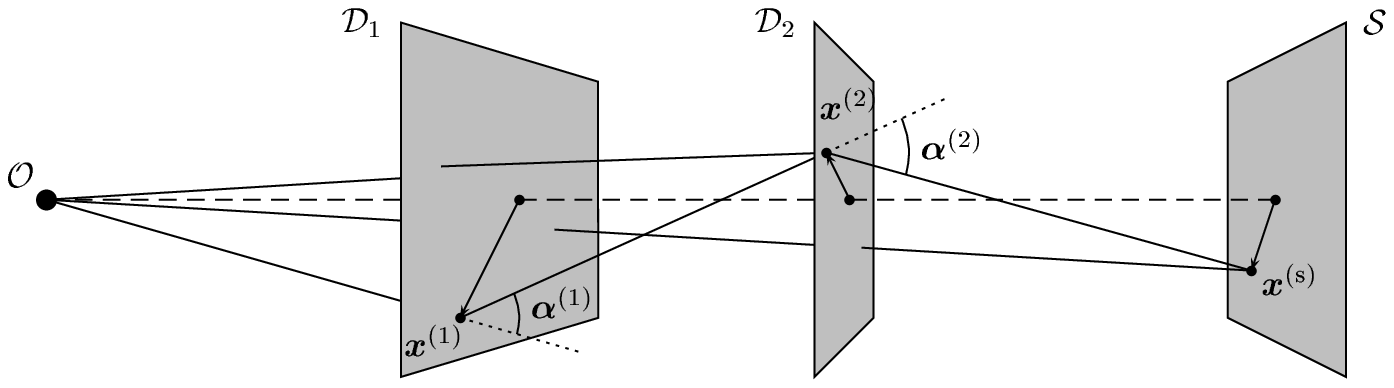}
\end{center}
\figcaption[f1.ps]{The geometrical configuration of a double lens
  system: the observer $\mathcal{O}$, the two deflector planes
  $\mathcal{D}_1$ and $\mathcal{D}_2$, and the source plane
  $\mathcal{S}$.\label{Fig:1}}

\begin{center}
  \includegraphics{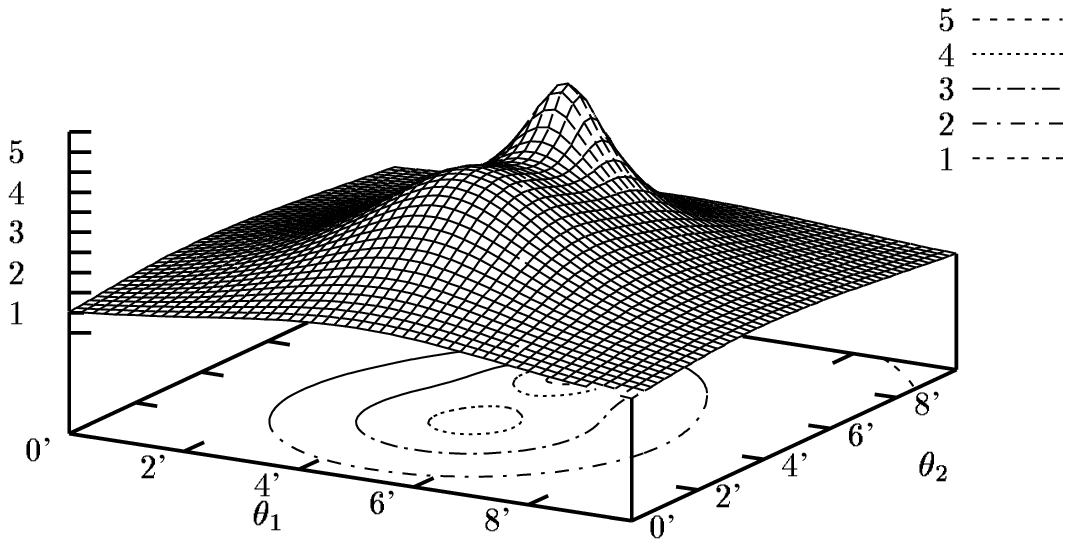}
\end{center}
\figcaption[f2.ps]{This plot shows the light distribution, in
  arbitrary units, that would be observed if the two clusters had
  equal and constant mass-to-light ratio. Note that the peak
  corresponding to the center of the far cluster is slightly higher,
  because of the contribution by the near cluster, which appears to be
  more diffuse.\label{Fig:light2}}

\begin{center}
  \includegraphics{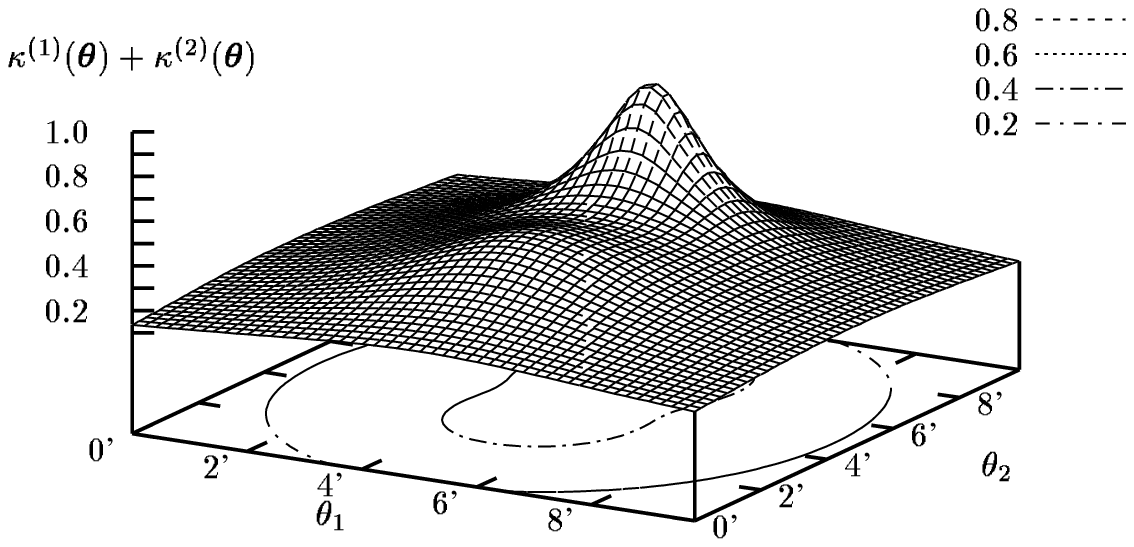}
\end{center}
\figcaption[f3.ps]{The sum of the dimensionless mass
  distributions $\kappa^{(1)}(\vec\theta)$ and
  $\kappa^{(2)}(\vec\theta)$ for the strong case. Note that, although
  the two clusters are physically identical, the near cluster appears
  to be significantly weaker and more diffuse. The maximum value
  reached by the combined dimensionless mass distribution is
  $\kappa^{(1)} + \kappa^{(2)} \simeq 1.072$. However, the lens is
  still sub-critical (the maximum value for $1 - \sigma$ is about
  $0.928$).\label{Fig:density2}}

\begin{center}
  \includegraphics{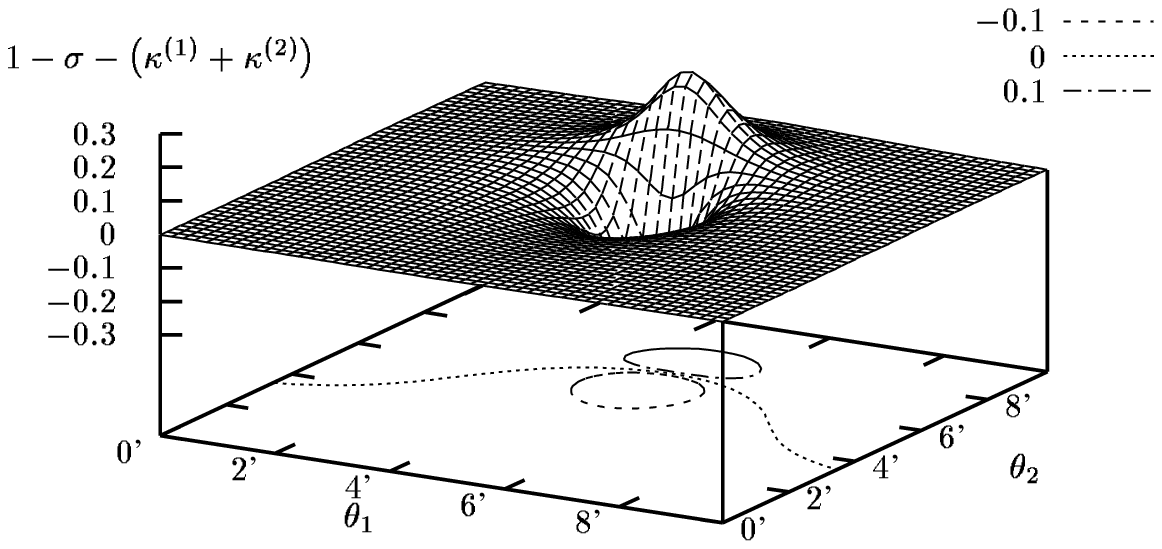}
\end{center}
\figcaption[f4.ps]{The quantity $1 - \sigma -
  \bigl(\kappa^{(1)} + \kappa^{(2)} \bigr)$ plotted here for the
  strong case shows a ``hole'' and a ``peak'' which can be understood
  in terms of the lens effect of the near cluster on the ``effective''
  mass of the second cluster. The quantity shown here vanishes only in
  the weak lensing limit.\label{Fig:trA0-density2}}

\begin{center}
  \includegraphics{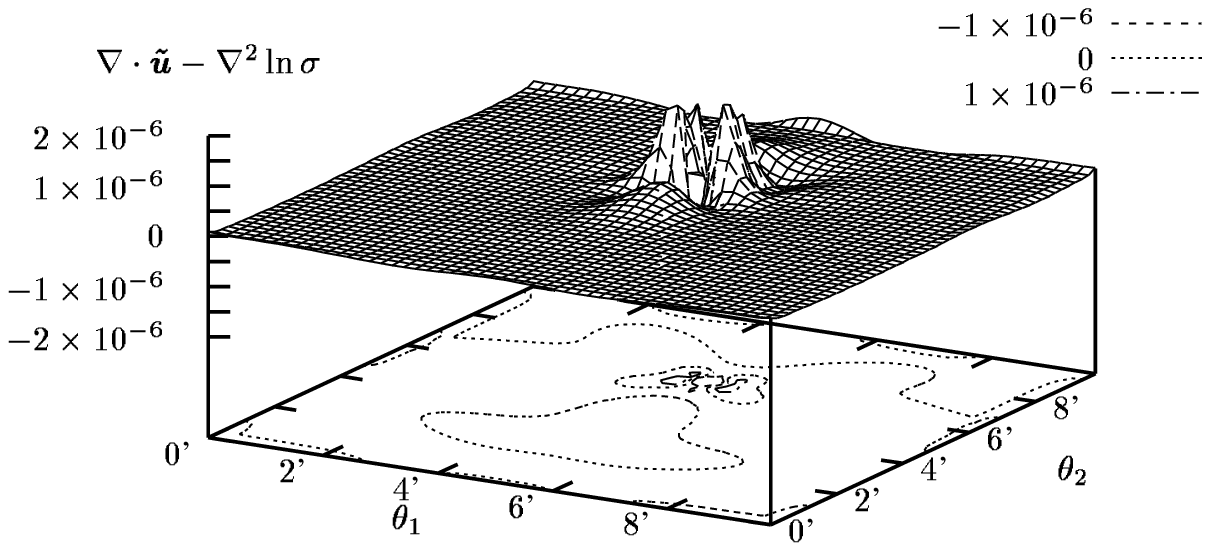}
\end{center}
\figcaption[f5.ps]{The difference $\nabla \cdot
  \vec{\tilde u} - \nabla^2 \ln \sigma$ on the $\vec\theta$-plane for
  the weak case. As described in the text, to second order this
  quantity is expected to vanish (Eq.~\eqref{eq:B26}). For comparison,
  we note that the maximum value of $|\nabla \cdot \vec{\tilde u}|$ is
  about $0.00647$.\label{Fig:divu-lapsigma1}}

\begin{center}
  \includegraphics{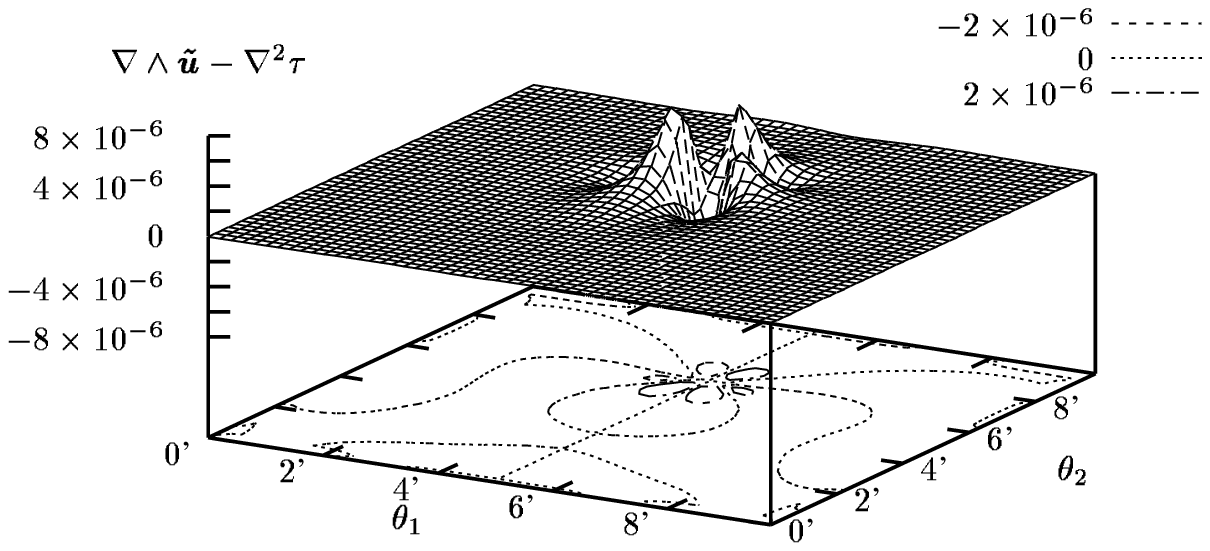}
\end{center}
\figcaption[f6.ps]{The difference $\nabla \wedge
  \vec{\tilde u} - \nabla^2 \tau$ on the $\vec\theta$-plane for the
  weak case. As described in the text, to second order this quantity
  is expected to vanish (Eq.~\eqref{eq:B27}). The maximum value of $|
  \nabla \wedge \vec{\tilde u}|$ is about $3.37 \times
  10^{-5}$.\label{Fig:curlu-laptau1}}

\begin{center}
  \includegraphics{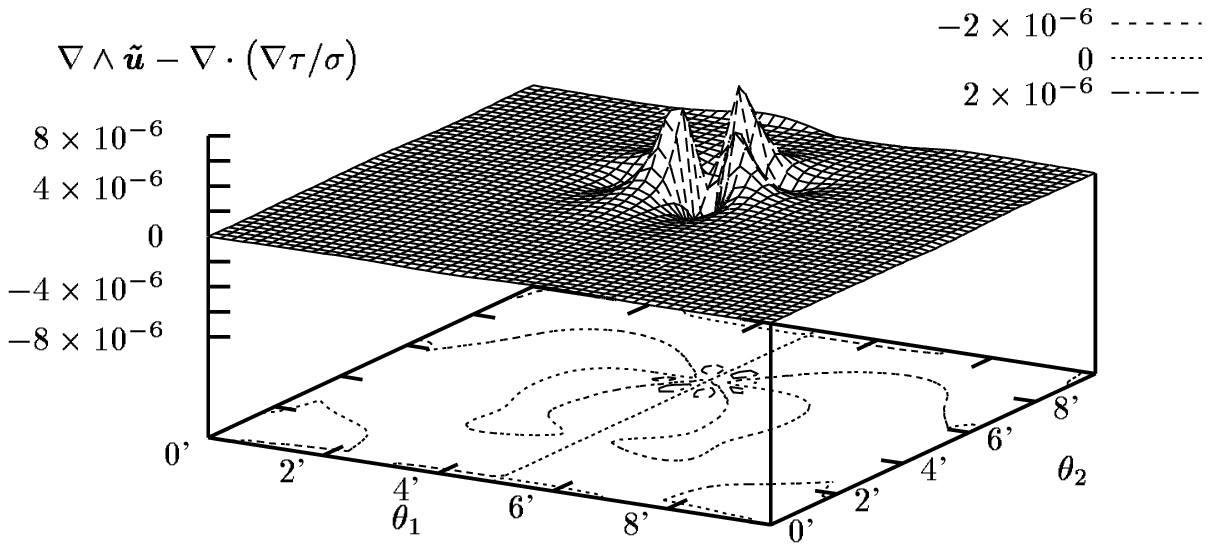}
\end{center}
\figcaption[f7.ps]{The difference $\nabla \wedge
  \vec{\tilde u} - \nabla \cdot (\nabla \tau / \sigma)$ on the
  $\vec\theta$-plane for the weak case. To second order, $\nabla \cdot
  (\nabla \tau / \sigma) = \nabla^2 \tau$, and in fact the plot shown
  here is very similar to that of
  Fig.~\ref{Fig:curlu-laptau1}.\label{Fig:curlu-laptau_1}}

\begin{center}
  \includegraphics{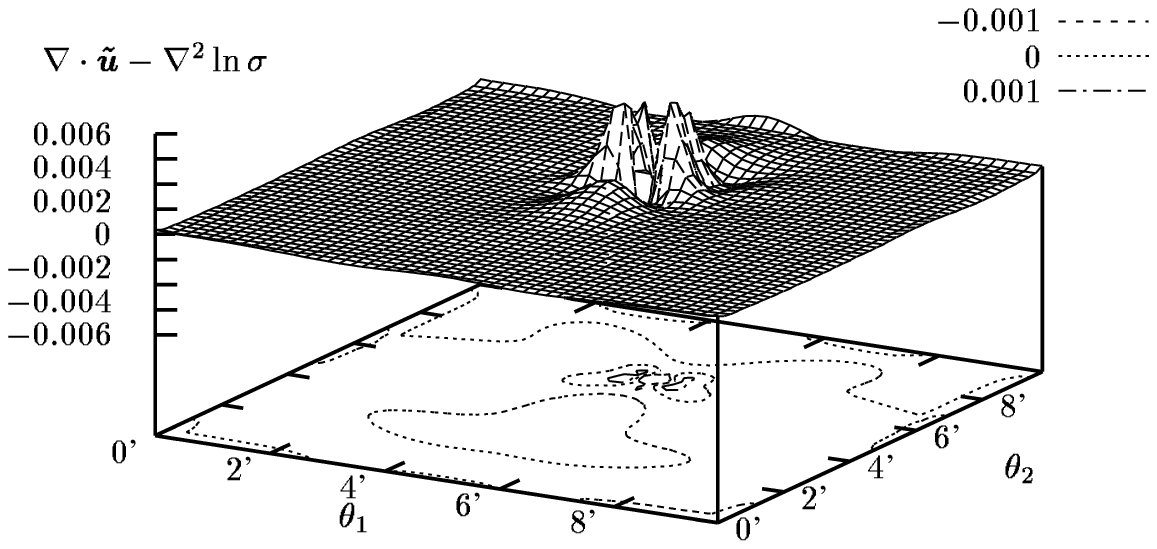}
\end{center}
\figcaption[f8.ps]{As in Fig.~\ref{Fig:divu-lapsigma1},
  but for the strong case. Note that the difference shown remains
  small. For comparison, the maximum value of $| \nabla \cdot
  \vec{\tilde u} |$ is about $0.331$. \label{Fig:divu-lapsigma2}}

\begin{center}
  \includegraphics{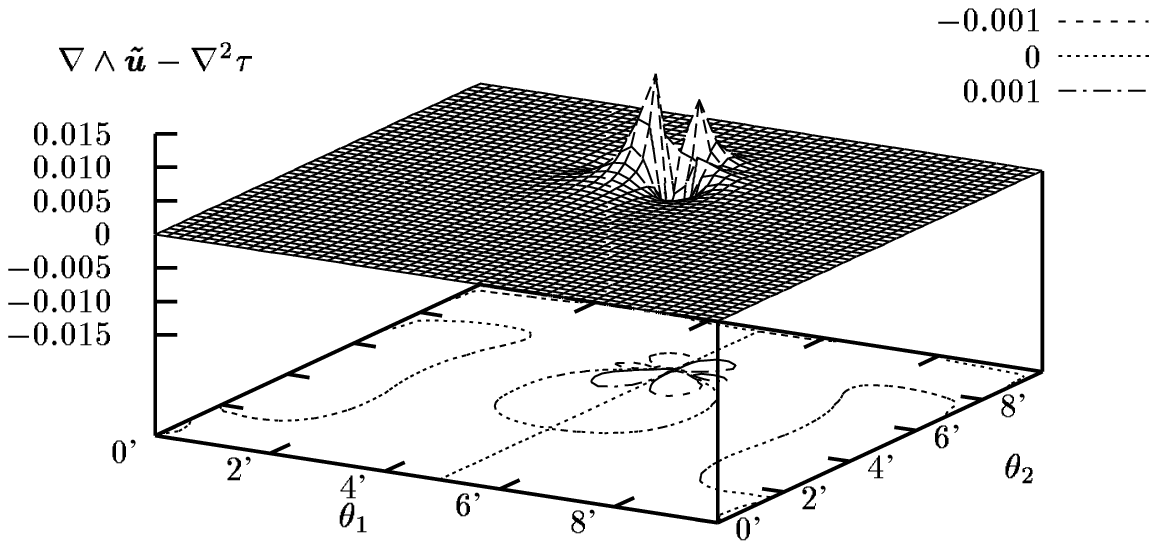}
\end{center}
\figcaption[f9.ps]{As in Fig.~\ref{Fig:curlu-laptau1}, but
  for the strong case. The maximum value of $| \nabla \wedge
  \vec{\tilde u} |$ is about $0.0167$. \label{Fig:curlu-laptau2}}

\begin{center}
  \includegraphics{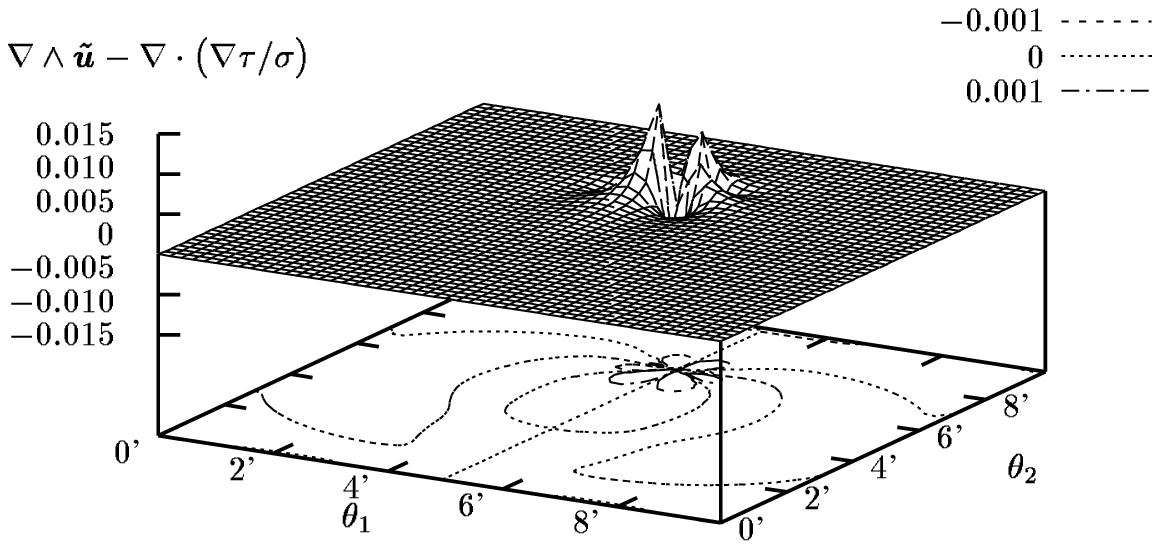}
\end{center}
\figcaption[f10.ps]{As in Fig.~\ref{Fig:curlu-laptau_1}, but
  for the strong case.\label{Fig:curlu-laptau_2}}

\begin{center}
  \includegraphics{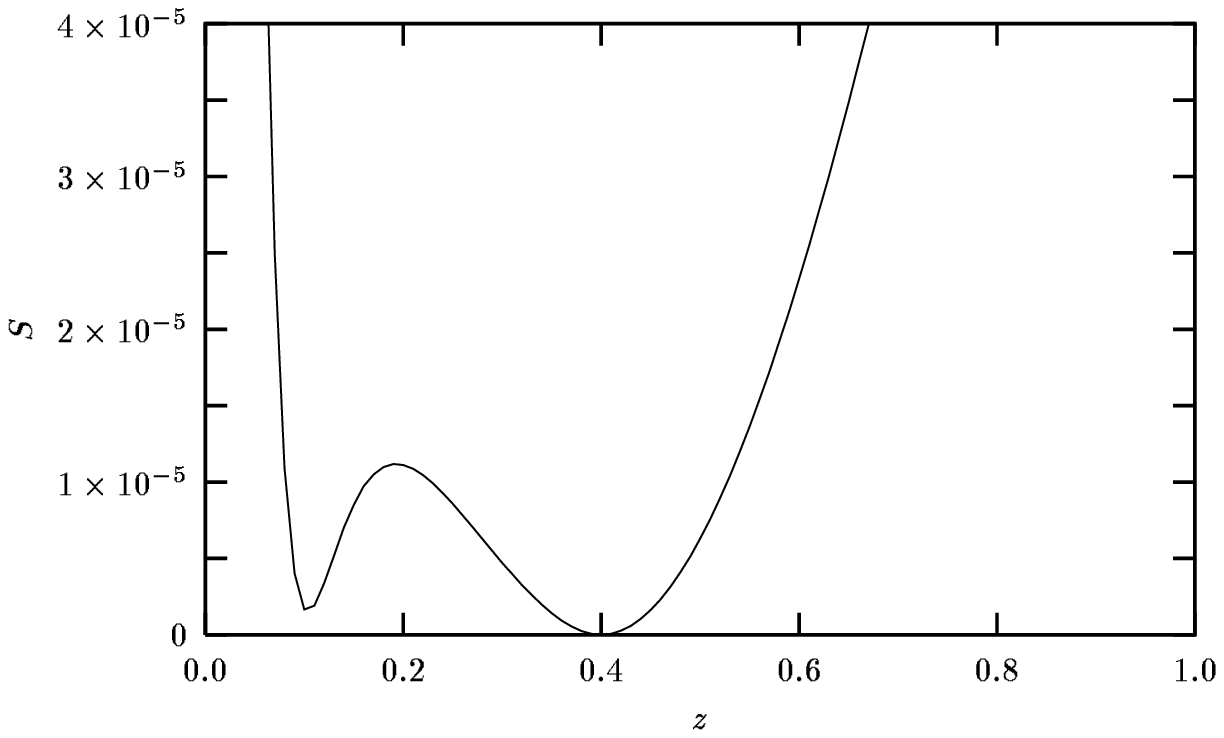}
\end{center}
\figcaption[f11.ps]{The detection of a dark cluster at redshift
  $z = 0.1$. The value of the integral $S$ (see text for definition)
  is shown as a function of the redshift $z$ of the dark cluster. Note
  that $S$ does not exactly vanish at $z = 0.1$ because of
  discretization errors.  The luminous cluster is at redshift
  $z_\mathrm{lum} = 0.4$.\label{Fig:findz-0.1}}

\begin{center}
  \includegraphics{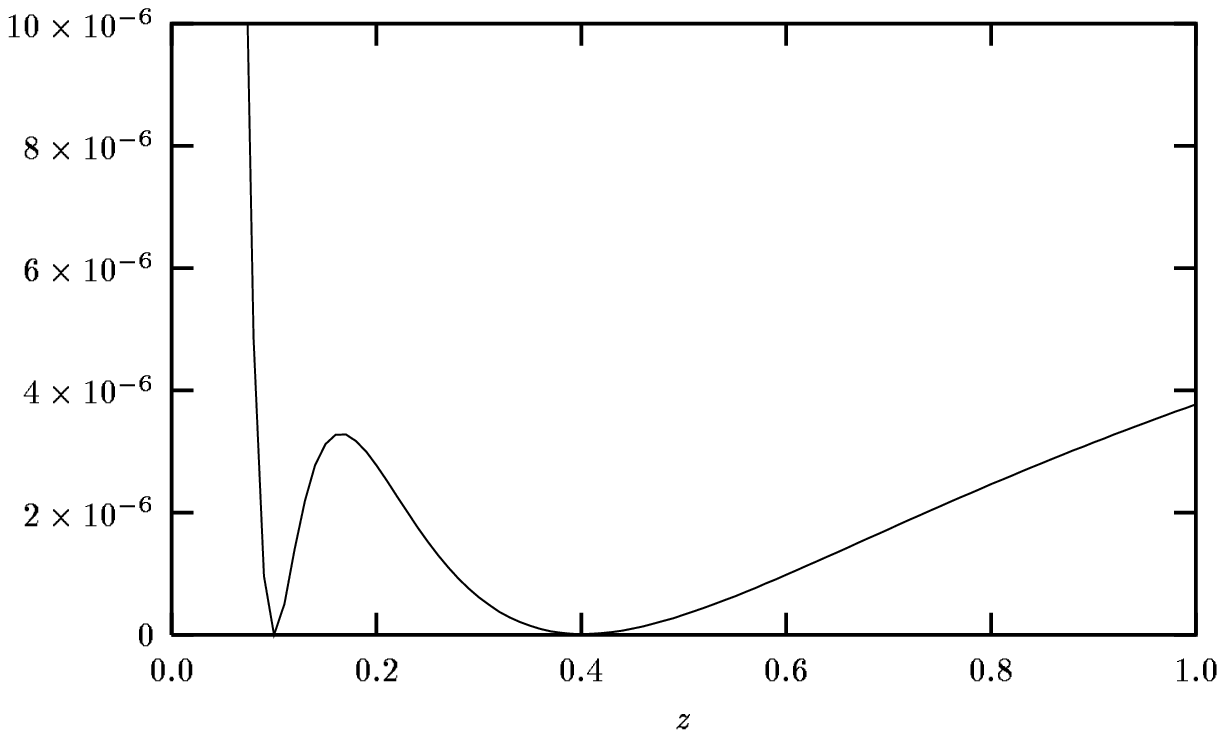}
\end{center}
\figcaption[f12.ps]{As for Fig.~\ref{Fig:findz-0.1}, but with
  the luminous and the dark cluster locations swapped. The function
  $S$ has been computed on sampled redshifts, and for this reason the
  minimum at $z=0.1$ appears to be somewhat jagged.\label{Fig:findz-0.4}}

\begin{center}
  \includegraphics{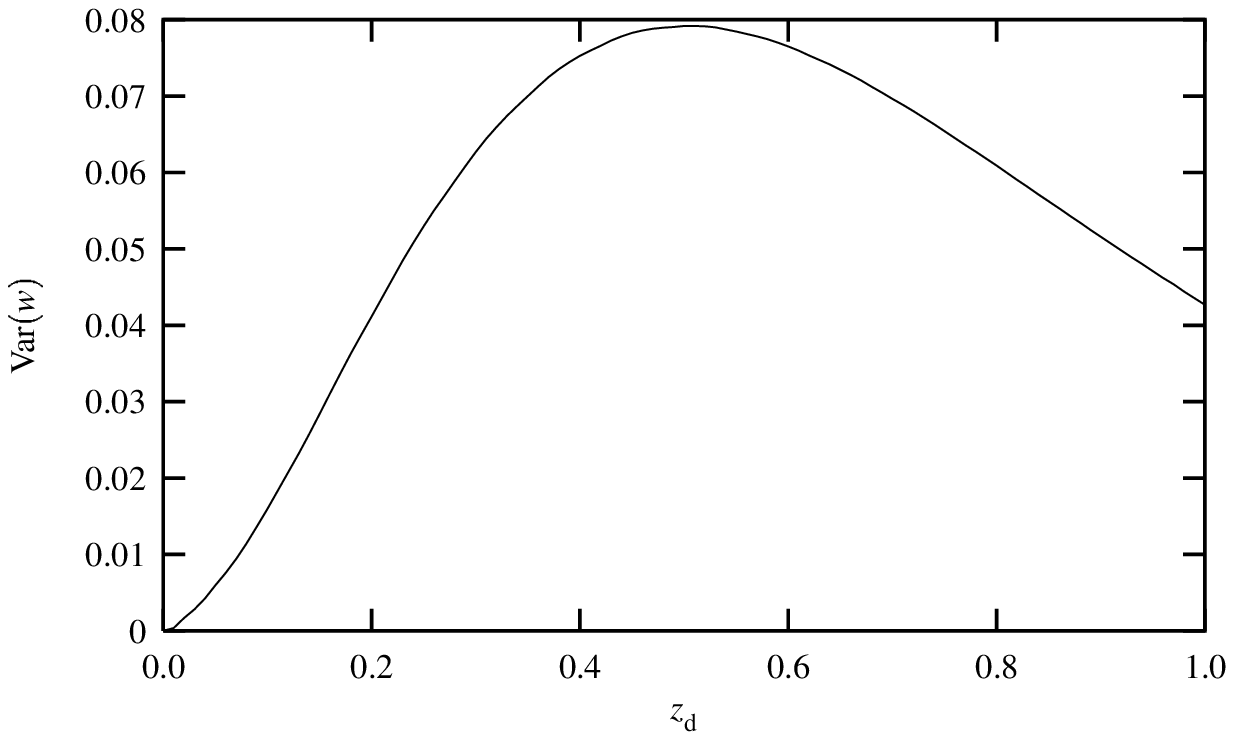}
\end{center}
\figcaption[f13.ps]{Variance of $w$ as a function of the lens
  redshift $z_\mathrm{d}$ ($\Omega_0 = 0.3, \Omega_\Lambda = 0.7$).
  The assumed source probability distribution is given by
  Eq.~\eqref{eq:14}. \label{Fig:var_wz}}

\begin{center}
  \includegraphics{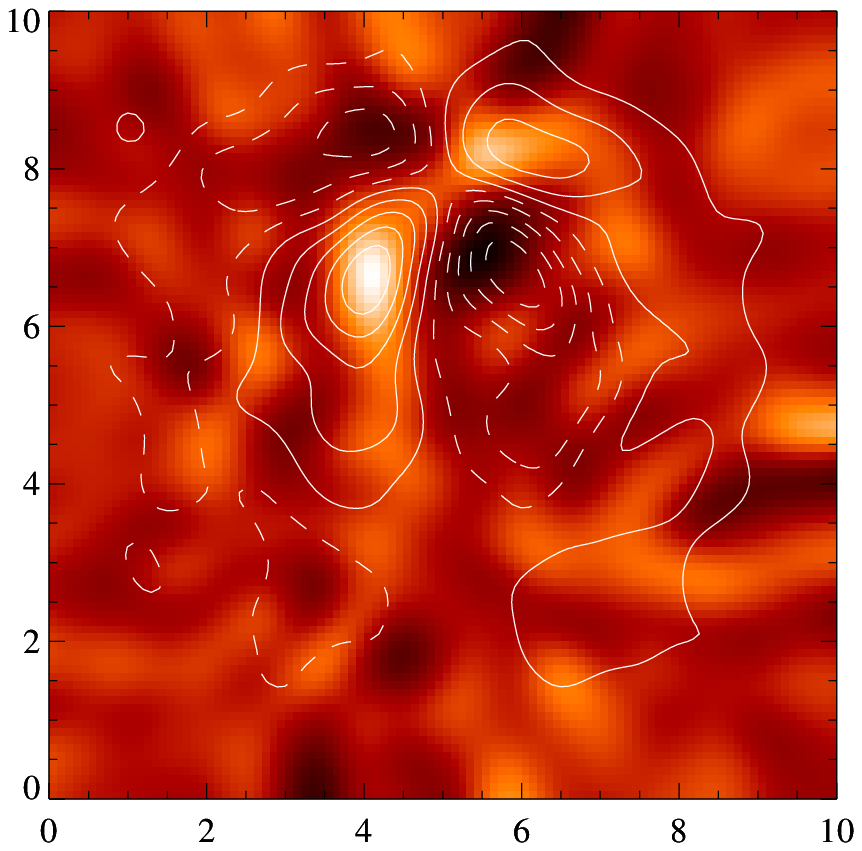}
\end{center}
\figcaption[f14.ps]{Density plot of $\nabla \wedge \vec {\tilde u}$
  for the ``strong'' double lens. This plot has been obtained by
  simulating the observation of galaxies in a field of $10' \times
  10'$. The adopted galaxy density is $100 \mbox{ gal arcmin}^{-2}$.
  The superimposed contour plot describes the quantity $\nabla \wedge
  \vec{\tilde u}_0$, after suitable smoothing. Contours are at levels
  $[\pm 3, \pm 9, \pm 15, \pm 21, \pm 27] \times 10^{-5}$ (positive
  contours are solid, negative ones are dashed). Because of a
  different choice for the unit of length, the contour levels of
  $\nabla \wedge \vec {\tilde u}$ shown here should not be compared
  directly with those plotted in Fig.~\ref{Fig:curlu-laptau2}. The
  plane represented in the figure is precisely the $(\theta_1,
  \theta_2)$ plane as drawn in Fig.~\ref{Fig:density2}.
  \label{Fig:dens_plot}}

\begin{center}
  \includegraphics{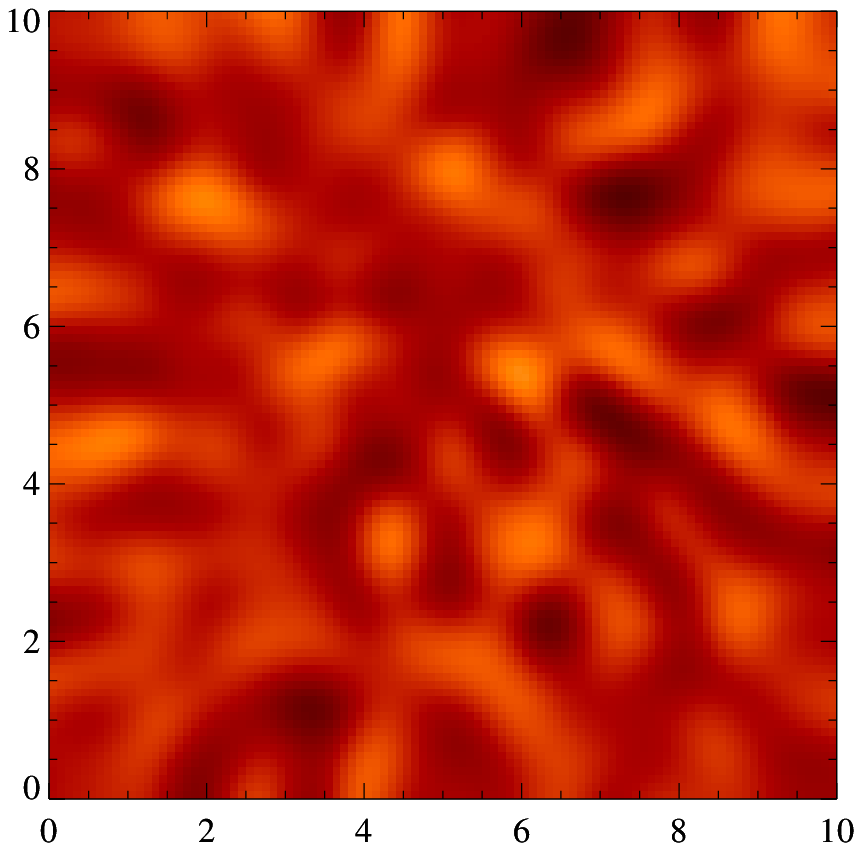}
\end{center}
\figcaption[f15.ps]{Same as Fig.~\ref{Fig:dens_plot} but for a
  single lens at redshift $z = 0.1$. Equal gray levels in the two
  pictures correspond to equal values for $\nabla \wedge \vec {\tilde
    u}$. A comparison of the signal in the previous figure with the
  noise observed here provides an estimate of the detection
  significance of a double lens. Note that most of the noise in this
  map is due to the finite ellipticity of source galaxies and, to a
  smaller extent, to the spread in redshift of galaxies.
  \label{Fig:density-single}}

\end{document}